\documentclass[journal,onecolumn,twoside]{IEEEtran}
\normalsize
\ifCLASSINFOpdf
\else
\fi
\usepackage{setspace}
\usepackage{color}
\usepackage{cite}
\usepackage{amsmath}
\usepackage{amsfonts}
\usepackage{amssymb}
\usepackage{amsthm}
\usepackage{tikz}
\usepackage{cases}
\usepackage{bm}
\usepackage{graphicx}
    \graphicspath{{../}}
    \DeclareGraphicsExtensions{.pdf}
\usepackage[caption=true,font=footnotesize]{subfig}
\usepackage{multirow}
\usepackage{makecell}
\usepackage{mathdots}
\usepackage{booktabs}
\usepackage{url}
\usepackage{bm}
\usepackage{xtab}
\usepackage{pifont}
\usepackage{array}
\usepackage{algorithm}
\usepackage{algorithmic}
\usepackage{enumerate}
\usetikzlibrary{arrows}

\usepackage[normalem]{ulem}
\usepackage{lineno}

\allowdisplaybreaks[4]
\usepackage[colorlinks,
            linkcolor=blue,
            anchorcolor=blue,
            citecolor=blue]{hyperref}
\theoremstyle{plain}

\newtheorem{theorem}{Theorem}
\newtheorem{lemma}[theorem]{Lemma}

\newtheorem{corollary}[theorem]{Corollary}
\theoremstyle{definition}
\newtheorem{example}{Example}

\newtheorem{remark}{Remark}

\begin{document}	

\title{New Constructions of Binary Cyclic Codes with Both Relatively Large
Minimum Distance and Dual Distance}
\author{Lingqi Zheng,~\IEEEmembership{}
        Weijun~Fang,~\IEEEmembership{}
        Rongxing Qiu~\IEEEmembership{}
\IEEEcompsocitemizethanks{\IEEEcompsocthanksitem Lingqi Zheng, Weijun Fang and Rongxing Qiu are with the State Key Laboratory of Cryptography and Digital Economy Security, the Key Laboratory of Cryptologic Technology and Information Security, Ministry of Education, and the School of Cyber Science and Technology, Shandong University, Qingdao 266237, China  (e-mail: zlingqihhh@outlook.com, fwj@sdu.edu.cn, qrx@mail.sdu.edu.cn).

Weijun Fang is also with  Quan Cheng Laboratory, Jinan 250103, China.}
\thanks{The work is supported in part by the National Key Research and Development Program of China under Grant Nos. 2021YFA1001000 and 2022YFA1004900, the National Natural Science Foundation of China under Grant No. 62571301, and the Shandong Provincial Natural Science Foundation under Grant No. ZR2025QA05. {\it (Corresponding Author: Weijun Fang)}.}
\thanks{Manuscript submitted }}


\maketitle
\begin{abstract}
Binary cyclic codes are worth studying due to their applications and theoretical importance. It is an important problem to construct an infinite family of cyclic codes with large minimum distance $d$ and dual distance $d^{\perp}$. In recent years, much research has been devoted to improving the lower bound on $d$, some of which have exceeded the square-root bound. The constructions presented recently seem to indicate that when the minimum distance increases, the minimum distance of its dual code decreases. In this paper, we focus on the new constructions of binary cyclic codes with length $n=2^m-1$, dimension near $n/2$ and both relatively large minimum distance and dual distance. When $m$ is even, we construct a family of binary cyclic codes with parameters $[2^m-1,2^{m-1}\pm1,d]$, where $d\ge 2^{m/2}-1$ and $d^\perp\ge2^{m/2}$. Both the minimum distance and the dual distance are significantly better than the previous results. When $m$ is the product of two distinct primes, we construct some cyclic codes with dimensions $k=(n+1)/2$ and $d>\frac{n}{\log_2n},$ where the lower bound on the minimum distance is much larger than the square-root bound. When $m$ is odd, we present two families of binary $[2^m-1,2^{m-1},d]$ cyclic codes with $d\ge2^{(m+1)/2}-1$, $d^\perp\ge2^{(m+1)/2}$ and $d\ge2^{(m+3)/2}-15$, $d^\perp\ge2^{(m-1)/2}$ respectively, which leads that $d\cdot d^\perp$ can reach $2n$ asymptotically. To the best of our knowledge, for the binary cyclic codes with length $n=2^m-1$ and dimension $k=(n\pm1)/2$, except for the punctured binary Reed-Muller codes, there is no other construction of binary cyclic codes that reaches this bound. 
\end{abstract}


\IEEEpeerreviewmaketitle

\section{Introduction}
\label{se1}
\IEEEPARstart{L}{et} $\mathbb{F}_2^n$ be the $n$-dimensional vector space over the finite field $\mathbb{F}_2=\left\{0,1\right\}$. The Hamming weight $wt(\bm{a})$ of a vector $\bm{a}=(a_0,a_1,...,a_{n-1})\in\mathbb{F}_2^n$ is the cardinality of the set $supp(\bm{a})=\left\{0\le i\le n-1:a_i=1\right\}.$ The Hamming distance $d(\bm{a},\bm{b})$ between $\bm{a}$ and $\bm{b}$ is $d(\bm{a},\bm{b})=wt(\bm{a}-\bm{b})$. For a code $C\subseteq \mathbb{F}_2^n$, the minimum Hamming distance of $C$ is defined as
$$d=\min_{\bm{a}\not=\bm{b}}\left\{d(\bm{a},\bm{b}):\bm{a}\in C,\bm{b}\in C\right\}.$$

An $[n,k,d]_2$ linear code $C$ is a $k$-dimensional linear subspace of $\mathbb{F}_2^n$ with minimum Hamming distance $d$. Let $\bm{u}=(u_0,u_1,...,u_{n-1})\in\mathbb{F}_2^n$ and $\bm{v}=(v_0,v_1,..,v_{n-1})\in\mathbb{F}_2^n$, the Euclidean inner product is defined as
$$\langle\bm{u},\bm{v}\rangle=\sum_{i=0}^{n-1}u_iv_i,$$
then the Euclidean dual code of $C$ is defined as 
$$C^\perp=\left\{\bm{x}\in\mathbb{F}_2^n:\langle\bm{x},\bm{c}\rangle=0,\forall\bm{c}\in C\right\}$$
and we denote the minimum distance of $C^\perp$ as $d^\perp$, which is called as the dual distance of $C$.

 Cyclic codes were introduced by E. Prange in \cite{ref15}. We say that an $[n,k]$ linear code $C$ is cyclic if and only if for any $\bm{c}=(c_0,c_1,...,c_{n-1})\in C$, we have $(c_{n-1},c_0,c_1,...,c_{n-2})\in C.$ Let $n$ be an odd integer and let $C$ be a binary cyclic code of length $n$. A codeword $\bm{c}=(c_0,c_1,...,c_{n-1})$ in $C$ can be identified with the polynomial
$$c(x)=c_0+c_1x+\cdots+c_{n-1}x^{n-1}\in\mathbb{F}_2[x]/\langle x^n-1\rangle.$$
It is not hard to prove that the linear code $C$  is cyclic if and only if the set $\left\{c(x)=c_0+c_1x+\cdots+c_{n-1}x^{n-1}:(c_0,c_1,...,c_{n-1})\right.$\\$\left.\in C\right\}$ is an ideal in $\mathbb{F}_2[x]/\langle x^n-1\rangle$.
Meanwhile, we know that every ideal of $\mathbb{F}_2[x]/\langle x^n-1\rangle$ is principal. Then there is a monic polynomial $g(x)\in\mathbb{F}_2[x]$ with the smallest degree among all generators of $C$. Clearly, $g(x)$ is unique and defined as the generator polynomial. Moreover, the dimension of $C$ ${\rm{dim}}(C)=n-{\rm{deg}}(g(x))$. Let $h(x)=\frac{x^n-1}{g(x)}$, then the dual code $C^\perp$ of $C$ is generated by 
$$g^\perp(x)=\frac{x^{{\rm{dim}}(C)}h(x^{-1})}{h(0)}.$$ Obviously, ${\rm{dim}}(C)+{\rm{dim}}(C^\perp)=n.$ 

To obtain cyclic codes with improved parameters, many researchers have directed their attention toward BCH codes. This is primarily because they constitute an essential subclass of cyclic codes with excellent algebraic properties, and many BCH codes are known to be optimal linear codes. Binary BCH codes were introduced by Hocquenghem \cite{ref24}, Bose and Ray-Chaudhuri \cite{ref20}. Let $m\ge 2$ be a positive integer and $n=2^m-1$. Let $\beta$ be a primitive element of $\mathbb{F}_{2^m}$. For any integer $i$, let ${\rm{m}}_{\beta^i}(x)$ denote the minimal polynomial of $\beta_i$ over $\mathbb{F}_2$, which is the monic polynomial of the smallest degree over $\mathbb{F}_2$ having $\beta^i$ as a root. For any integer $2\le\delta\le n$, define
$$g_{(m,b,\delta)}(x)={\rm{lcm}}({\rm{m}}_{\beta^b}(x),{\rm{m}}_{\beta^{b+1}}(x),...,{\rm{m}}_{\beta^{b+\delta-2}}(x)),$$
 where $b$ is an integer and ${\rm{lcm}}(\cdot)$ denotes the least common multiple of the minimal polynomials ${\rm{m}}_{\beta^i}(x)$ with $b \leq i \leq b+\delta-2.$ Let $C_{(m,b,\delta)}$ with generator polynomial $g_{(m,b,\delta)}(x)$ denote the binary BCH code with designed distance $\delta$. So far, much research has been done on the dimension of BCH codes with some special designed distances(\cite{ref26}, \cite{ref25}, \cite{ref27}), and the parameters of BCH codes with some special length have also been studied thoroughly(\cite{ref31}, \cite{ref26}, \cite{ref30},  \cite{ref32}, \cite{ref28}, \cite{ref10}, \cite{ref33}, \cite{ref7}). However, there are many kinds of BCH codes that still remain to be investigated.

\subsection{Related work}
Binary cyclic codes have been widely studied for 70 years due to their theoretical importance and practical applications. Many good properties and constructions of binary cyclic codes have been proposed during this time(\cite{ref34}, \cite{ref35}, \cite{ref36},\cite{ref37},\cite{ref38},\cite{ref39},\cite{ref40}).
We say the binary cyclic code is simple-root cyclic if $n$ is odd and repeated-root cyclic if $2|n$. When $n$ is odd, constructing an infinite family of binary $[n, (n\pm1)/2, d]$ cyclic codes with large minimum distance $d$ remains an important and challenging problem. Furthermore, designing a family of binary codes with both a good minimum distance $d$ and a good dual minimum distance $d^\perp$ presents an even greater challenge. To the best of our knowledge, the known infinite families of binary cyclic codes with large minimum distances can be summarized as follows:
\begin{itemize}
    \item Binary quadratic residue codes with parameters $[n,(n+1)/2,d]$ and their even-weight subcodes with parameters $[n,(n-1)/2,d+1]$, where $d^2\ge n$ and $n\equiv\pm1\pmod8$ is a prime \cite{ref11}. The lower bound of $d$ is defined as the square-root bound.
    
    \item The punctured binary Reed-Muller codes of order $(m-1)/2$ with parameters $[2^m-1,2^{m-1},2^{(m+1)/2}-1]$ and $d^\perp\ge 2^{(m+1)/2}$, where $m$ is odd \cite{ref8}.

    \item Two families of binary cyclic codes with parameters $[n_1n_2,(n_1n_2+1)/2,d],$ where $d>\max\left\{\sqrt{n_1},\sqrt{n_2}\right\}$, $n_1$ and $n_2$ are two distinct primes such that $n_1\equiv\pm1\pmod8$ and $n_2\equiv\pm1\pmod8$, or $n_1\equiv\pm3\pmod8$ and $n_2\equiv\pm3\pmod8$ \cite{ref12},\cite{ref13},\cite{ref14}.

    \item A family of binary cyclic codes with parameters$[2^m-1,2^{m-1},d],$ where $m\ge3$ is odd. If $m\equiv3\pmod4$, $d\ge2^{(m-1)/2}+1$ and $d^\perp\ge2^{(m-1)/2}+2$. If $m\equiv1\pmod4$, $d\ge2^{(m-1)/2}+3$ and $d^\perp\ge2^{(m-1)/2}+4$ \cite{ref4}.

    \item A family of binary cyclic codes with parameters $[2^m-1,2^{m-1}-1,d]$, where $m\ge4$ is even. If $m\equiv0\pmod4$, $d\ge2^{(m-2)/2}+1$ and $d^\perp\ge2^{(m-2)/2}+2$. If $m\equiv2\pmod4$, $d\ge2^{(m-4)/2}+1$ and $d^\perp\ge2^{(m-4)/2}+2$ \cite{ref4}.
    
    \item A family of binary cyclic codes with parameters $[2^m-1,2^{m-1},d]$, where $m\equiv1\pmod4\ge5$. If $m\equiv1\pmod8$, $d\ge2^{(m-1)/2}+3$ and $d^\perp\ge2^{(m-1)/2}+4$. If $m\equiv5\pmod8$, $d\ge2^{(m-1)/2}+1$ and $d^\perp\ge2^{(m-1)/2}+2$ \cite{ref6}.

    \item A family of binary cyclic codes with parameters $[2^m-1,2^{m-1},d]$, where $m\equiv3\pmod4$. If $m\equiv3\pmod8$, $d\ge 2^{(m-1)/2}+1$ and $d^\perp\ge2^{(m-1)/2}+2$. If $m\equiv7\pmod8$, $d\ge 2^{(m-1)/2}+3$ and $d^\perp\ge2^{(m-1)/2}+4$ \cite{ref6}.

    \item A family of binary cyclic codes with parameters $[2^m-1,2^{m-1},d],$ where $m\ge9$ is odd $d\ge3\cdot2^{(m-1)/2}-1$ and $d^\perp\ge 2^{\frac{m-1}{2}}$ \cite{ref5}.

    \item A family of binary cyclic codes with parameters $[2^m-1,2^{m-1},d]$, where $m$ is an odd prime, $d\ge\frac{2^m-1}{m}$ and $d^\perp\ge m$ \cite{ref7}.

    \item A family of binary cyclic codes with parameters $[n,\ge\frac{n+1}{2},d]$ with length $n=\sum_{i=1}^s(2^{p_i}-1)$ and $d\ge\lceil\frac{n-1}{\prod_{i=1}^sp_i}\rceil,$ where $p_1,p_2,...,p_s$ are different primes \cite{ref9}. However, the value of $d^\perp$ has not been studied.
\end{itemize}
It is an interesting problem to construct new infinite families of binary cyclic codes with length $n$, dimension close to $n/2$ and both relatively large minimum distance 
$d$ and dual distance $d^{\perp}$.  The following are some known constructions of binary cyclic codes with good parameters in this case. 

When $m$ is even, the construction of binary cyclic codes with parameters $[2^m-1, 2^{m-1}-1, d]$ remains underexplored. In \cite{ref4}, it is shown that the lower bound of the minimum distance is $2^{(m-2)/2} + 1$ for $m \equiv 0 \pmod{4}$ and $2^{(m-4)/2} + 1$ for $m \equiv 2 \pmod{4}$ respectively. However, these bounds are smaller than the square-root bound. In \cite{ref5}, Sun et al. constructed a family of binary cyclic codes with parameters $[2^m-1,2^{m-1}+1-\frac{m}{2},d\ge3\cdot2^{\frac{m-2}{2}}+1]$, whose dimension is slightly smaller than $2^{m-1}-1$ as $m$ increases. 

When $m$ is odd, the punctured Reed-Muller codes have parameters $[2^m-1, 2^{m-1}, d=2^{\frac{m+1}{2}}-1]$, in 
\cite{ref5}, Sun et al. gave a construction of binary cyclic codes with parameters $[2^m-1,2^{m-1},d\ge 3\cdot 2^{\frac{m-1}{2}}-1]$.
In addition, when $m$ is an odd prime, in \cite{ref7}, Sun analyzed a family of binary cyclic codes with parameters $[2^m-1,2^{m-1},d\ge\frac{2^m-1}{m}]$, whose minimum distance has lower bound about $\frac{n}{\log_2 n}$, which is significantly larger than the square-root bound.

\subsection{Our results}
In this paper, we focus on the new constructions of binary cyclic codes with length $n=2^m-1$, dimension close to $n/2$ and both relatively large minimum distance 
$d$ and dual distance $d^{\perp}$.  We summarize our main results as follows:
\begin{itemize}
    \item For even $m$, we present a new construction of binary cyclic codes with parameters $[2^m-1, 2^{m-1} - 1, d]$, where $d \ge 2^{\frac{m}{2}} - 1$ and $d^\perp \ge 2^{\frac{m}{2}}$. Compared to the results in \cite{ref4}, our construction achieves improved parameters, with both larger minimum distance and dual minimum distance.
    \item For $m=p_1 p_2$, where $p_1$ and $p_2$ are two different primes, 
    specifically, we present two constructions of cyclic codes with parameters: $\left[2^m-1,2^{m-1},d\ge \left\lfloor\frac{2^{p_1p_2}-1}{p_1p_2}\right\rfloor+\left\lfloor\frac{2^{p_1}-1}{2p_1p_2}\right\rfloor+\left\lfloor\frac{2^{p_2}-1}{2p_1p_2}\right\rfloor+\left\lfloor\frac{2^{p_1}-1}{4p_1p_2}\right\rfloor+\left\lfloor\frac{2^{p_2}-1}{4p_1p_2}\right\rfloor\right]$ when 
    $p_2>p_1\geq 3$ and $\left[2^m-1,2^{m-1},d\ge\frac{2^{2p_2}-4}{2p_2}+\frac{2^{p_2}-2p_2-2}{4p_2}+1\right]$ when $p_1=2$, respectively. In both cases, the minimum distance $d$ is much larger than $\frac{n}{\log_2 n}.$
    \item For odd $m$, we present two families of binary $[2^m-1,2^{m-1},d]$ cyclic codes with $d\ge2^{(m+1)/2}-1$, $d^\perp\ge2^{(m+1)/2}$ and $d\ge2^{(m+3)/2}-15$, $d^\perp\ge2^{(m-1)/2}$ respectively.  Both the minimum distance and the dual distance of the first family of our codes are no less than those of the punctured Reed-Muller codes.  Compared with the binary cyclic codes constructed in \cite{ref5}, which have parameters $[2^m-1,2^{m-1},d\ge3\cdot2^{(m-1)/2}-1]$ and $d^\perp\ge2^{(m-1)/2}$, the second family of our codes has larger minimum distance while maintaining the same dual distance.
\end{itemize}
Furthermore, recent research has shown that as the minimum distance $d$ of a cyclic code $C$ increases, the value of $d^\perp$ decreases accordingly. The known specific values of $d$ and $d^\perp$ for binary cyclic codes of length $n = 2^m - 1$ are listed in Tables \ref{table2} and \ref{table1}.In the fifth column of each table, $\approx cn$ means that the lower bound on $d \cdot d^\perp$ is of the form $cn + o(n)$ as $n$ tends to infinity. Thus, whether we can construct some cyclic codes with both large $d$ and $d^\perp$ seems to be difficult. In this paper, we construct several families of cyclic codes for which the lower bounds of $d\cdot d^\perp$ approach $n$ in Theorems \ref{thm4}, \ref{thm1} and \ref{thm5}. Moreover, we present two families of binary cyclic codes with $d\cdot d^\perp\approx2n$ in Section \ref{set3}. To the best of our knowledge, for the binary cyclic codes with length $n=2^m-1$ and dimension $k=(n\pm1)/2$, except for the punctured binary Reed-Muller codes, there is no other construction of binary cyclic codes that reaches this bound.

 \begin{table*}
            \renewcommand{\arraystretch}{1.5}
            \centering
            \caption{ Binary $[n,k]$ cyclic codes with minimum distance $d$ and dual distance $d^{\perp}$ when $m$ is even\centering}
            \label{table2}
            \begin{tabular}{|c|c|c|c|c|c|}
             \hline
                $m$&$k$& $d\ge$  & $ d^\perp\ge$  & $ d\cdot d^\perp$  &    Reference\\
                \hline
                $m\equiv4\pmod 8$&$2^{m-1}+1$&$2^{(m-4)/2}+1$&$2^{(m-4)/2}+2$&$\approx\frac{1}{16}n$&\cite{ref4}\\
                \hline
                $m\equiv2\pmod 4$&$2^{m-1}+1$&$2^{(m-2)/2}+1$&$2^{(m-2)/2}+2$&$\approx\frac{1}{4}n$&\cite{ref4}\\
                \hline
                $m\equiv0\pmod 4$&$2^{m-1}-1$&$2^{(m-2)/2}+1$&$2^{(m-2)/2}+2$&$\approx\frac{1}{4}n$&\cite{ref4}\\
                \hline
                $m\equiv2\pmod 4$&$2^{m-1}-1$&$2^{(m-4)/2}+1$&$2^{(m-4)/2}+2$&$\approx\frac{1}{16}n$&\cite{ref4}\\
                \hline
                $m$ is even&$2^{m-1}-1$&$2^{\frac{m}{2}}-1$& $2^{\frac{m}{2}}$ & $\approx n$ &Theorem \ref{thm4} \\
                \hline
             $m$ is even&$2^{m-1}+1$&$2^{\frac{m}{2}}-1$& $2^{\frac{m}{2}}$ & $\approx n$ &Theorem \ref{thm4} \\
             \hline
             $m=2p$ with odd prime $p$&$2^{m-1}$&$\frac{2^{2p}-4}{2p}+\frac{2^p-2p-2}{4p}+1$&$2^{\lfloor\log_22p\rfloor+1}-1$&$\approx n$&Theorem \ref{thm1}\\
             \hline
 \end{tabular}
        \end{table*}

        \begin{table*}[ht]
            \renewcommand{\arraystretch}{1.5}
            \centering
            \caption{Binary $[n,k]$ cyclic codes with minimum distance $d$ and dual distance $d^{\perp}$ when $m$ is odd\centering}
           \label{table1}
            \begin{tabular}{|c|c|c|c|c|c|}
                \hline
                $m$&$k$& $d\ge$  & $d^\perp\ge$  & $d\cdot d^\perp$  &    Reference\\
                \hline
                   $m\equiv1\pmod2$ &$2^{m-1}$&$d(PRM) = 2^{\frac{m+1}{2}}-1$  & $2^{\frac{m+1}{2}}$ & $\approx2n$ & \cite{ref8}\\
                \hline
                $m\equiv1\pmod2$&$2^{m-1}$&$3\cdot 2^{\frac{m-1}{2}} - 1$ & $2^{\frac{m-1}{2}} $ & $\approx\frac{3}{2}n$ &  \cite{ref5} \\
                \hline
                $m\equiv1\pmod2$&$2^{m-1}$&$2^{\frac{m+1}{2}}-1$&$ 2^{\frac{m+1}{2}}$& $\approx2n$ & Theorem \ref{tm3}\\
                \hline
                $m\equiv1\pmod2$&$2^{m-1}$&$4\cdot(2^{\frac{m-1}{2}}-4)+1$&$2^{\frac{m-1}{2}}$&$\approx2n$&Theorem \ref{th6}\\
              \hline 
                $m\equiv 3 \pmod 4$&$2^{m-1}$&$2^{\frac{m-1}{2}} +1$    &  $2^{\frac{m-1}{2}} +2$    &  $\approx\frac{1}{2}n$   &  \cite{ref4}       \\ 
                \hline
                $m\equiv 1 \pmod 4$&$2^{m-1}$&$2^{\frac{m-1}{2}} +3$  &  $2^{\frac{m-1}{2}} +4$    &  $\approx\frac{1}{2}n$   &  \cite{ref4}       \\ 
                \hline
                $m\equiv 1 \pmod 8$&$2^{m-1}$&$2^{\frac{m-1}{2}} +3$  &  $2^{\frac{m-1}{2}} +4$   & $\approx\frac{1}{2}n$ & \cite{ref6} \\
                \hline
                $m\equiv 5 \pmod 8$&$2^{m-1}$&$2^{\frac{m-1}{2}} +1$   &  $2^{\frac{m-1}{2}} +2$  & $\approx\frac{1}{2}n$ & \cite{ref6} \\
                \hline
                $m\equiv 3 \pmod 8$&$2^{m-1}$&$2^{\frac{m-1}{2}} +1$    &  $2^{\frac{m-1}{2}} +2$  & $\approx\frac{1}{2}n$ & \cite{ref6} \\
                \hline
               $m\equiv 7\pmod 8$&$2^{m-1}$& $2^{\frac{m-1}{2}} +3$    &  $2^{\frac{m-1}{2}} +4$   & $\approx\frac{1}{2}n$ & \cite{ref6} \\
                \hline
                 $m$ is prime&$2^{m-1}$&$\frac{2^m-2}{m}$  & $ m$ &$\approx n$ & \cite{ref7}\\
                               \hline
                $m=p_1p_2$ with odd primes $p_1<p_2$ &$2^{m-1}$&
                $\substack{\left\lfloor\frac{2^{p_1p_2}-1}{p_1p_2}\right\rfloor+\left\lfloor\frac{2^{p_1}-1}{2p_1p_2}\right\rfloor\\
                +\left\lfloor\frac{2^{p_2}-1}{2p_1p_2}\right\rfloor+\left\lfloor\frac{2^{p_1}-1}{4p_1p_2}\right\rfloor+\left\lfloor\frac{2^{p_2}-1}{4p_1p_2}\right\rfloor}$&$2^{\lfloor\log_2p_1p_2\rfloor+1}$&$\approx n$&Theorem \ref{thm5}\\
                \hline
            \end{tabular}
        \end{table*}    
        \subsection{Organization}
For the rest of this paper, we present several families of binary cyclic codes of length $n=2^m-1$ and dimension $k=(n\pm1)/2$. Section \ref{set2} introduces several basic definitions and results. In Section \ref{set00}, we construct an infinite family of binary cyclic codes for even $m$. Section \ref{sect4} investigates the case where $m=p_1p_2$, with $p_1$ and $p_2$ being distinct primes. Section \ref{set3} studies the case where $m$ is odd. Finally, Section \ref{set6} concludes this paper and proposes an open problem regarding the values of $d \cdot d^\perp$.

\section{Preliminaries}
\label{set2}

In this section, we provide some preliminaries that will be used in the subsequent sections. Throughout this section, denote ${\rm{ord}}_n(i)$ as the smallest positive integer such that $i^{{\rm{ord}}_n(i)} \equiv 1 \pmod n$. For a binary linear code $C$ of length $n$, we define ${\rm{dim}}(C)$ and $d(C)$ as its dimension and minimum distance, respectively.

Let $\mathbb{Z}_n=\left\{0,1,...,n-1\right\}$ be the ring of integers modulo $n$. For any $i\in\mathbb{Z}_n$, the 2-cyclotomic coset $C_i^{(2,n)}$ modulo $n$ of $i$ is defined as
$$C_i^{(2,n)}=\left\{i2^j (\bmod~ n)|0\le j\le \ell_i-1\right\},$$
where $\ell_i$ is the smallest positive integer $\ell$ such that $i2^{\ell}\equiv i\pmod n$, hence $|C_i^{(2,n)}|=\ell_i.$ We define the smallest integer in $C_i^{(2,n)}$ as the coset leader of $C_i^{(2,n)}$. Let $\Gamma_{(2,n)}$ be the set of all coset leaders, then we have
$$\mathbb{Z}_n=\bigcup_{i\in\Gamma_{(2,n)}}C_i^{(2,n)}$$
and for any distinct elements $a,b\in\Gamma_{(2,n)}$, we have
$$C_a^{(2,n)}\bigcap C_b^{(2,n)}=\emptyset.$$

Let $m={\rm{ord}}_n(2)$ and let $\beta$ be a primitive $n$-th root of unity in $\mathbb{F}_{2^m}$. It is obvious that 
$$x^n-1=\prod_{i\in \mathbb{Z}_n}(x-\beta^i)=\prod_{i\in\Gamma_{(2,n)}}{\rm{m}}_{\beta^i}(x),$$
where $${\rm{m}}_{\beta^i}(x)=\prod_{j\in C_i^{(2,n)}}(x-\beta^j)\in\mathbb{F}_2[x].$$
Then ${\rm{m}}_{\beta^i}(x)$ is the minimal polynomial of $\beta^i.$

Let $C$ be a binary cyclic code of length $n$ with generator polynomial $g(x)$.
Then there exists a subset $T\subseteq\Gamma_{(2,n)}$ such that
$$g(x)=\prod_{i\in T}{\rm{m}}_{\beta^i}(x).$$
We define the set $Z = \cup_{i\in T} C_i^{(2,n)}$ as the defining set of $C$ with respect to $\beta$. Then ${\rm{dim}}(C)=n-|Z|$.  From the generator polynomial of $C^{\perp}$, we know that the defining set of $C^{\perp}$ is given by $-(\mathbb{Z}_n\backslash Z)=\{-i: i \in \mathbb{Z}_n\backslash Z\}$. The following is the well-known BCH bound, a fundamental result for cyclic codes.
\begin{lemma}[\hspace{-0.01em}\cite{ref1}]
    Let $C$ be a binary cyclic
code of length $n$ with a defining set $Z$. If there are integers $\delta$ and $h$ with $2\le \delta\le n$ such that $\left\{h+i\pmod n:0\le i\le\delta-2\right\}\subseteq Z$, then the minimum distance $d(C)$ of the code $C$ is at least $\delta$.
\end{lemma}

Let $m$ be a positive integer. In this paper, we always let $n=2^m-1$.
Then for the set 
$$[0,n]=\left\{0,1,2,...,2^m-1\right\},$$ define
\begin{align*}
    \pi: [0,n]&\rightarrow \mathbb{F}_2^m\\
    x&\mapsto (x_0,x_1,...,x_{m-1}),
\end{align*}
where $x=\sum_{i=0}^{m-1}x_i2^i.$ Clearly, $\pi$ is a bijection. We define the cyclic shifts of $(x_0,x_1,...,x_{m-1})$ as $$\rho(x_0,x_1,..,x_{m-1})=(x_{m-1},x_0,x_1,...,x_{m-2}).$$ 
The following lemma provides an easy criterion for a subset $A$ of $\mathbb{Z}_n$ to be a union of 2-cyclotomic sets under the action of the map $\pi$.
\begin{lemma}\label{lem1}
    Let $m$ be a positive integer and $n=2^m-1$. Then a subset $A$ of $\mathbb{Z}_n$ is a union of 2-cyclotomic sets of $\mathbb{Z}_n$ if and only if $\rho(\pi(A))=\pi(A).$
\end{lemma}
\begin{IEEEproof}
For any $x\in A$, let $\pi(x)=(x_0,x_1,...,x_{m-1})\in\mathbb{F}_2^m$, where $x=\sum^{m-1}_{i=0}x_i2^i$. Then
 $\rho(\pi(x))=(x_{m-1},x_0,x_1,...,x_{m-2}).$
 It leads to 
        $$\pi^{-1}(\rho(\pi(x)))=x_{m-1}+\sum_{i=1}^{m-1}x_{i-1}2^i\equiv2x\pmod{n}.$$
  Suppose $A$ is a union of some 2-cyclotomic sets of $\mathbb{Z}_n$, then for any $x\in A$, we have $\pi^{-1}(\rho(\pi(x)))\in C_x^{(2,n)}\subseteq A.$ It leads to $\rho(\pi(x))\in\pi(A)$, then $\rho(\pi(A)) \subseteq \pi(A).$ Since $\rho$ is a bijection, we can conclude that $\rho(\pi(A))=\pi(A).$
      Conversely, if $\rho(\pi(A))=\pi(A)$, then $\pi^{-1}(\rho(\pi(A)))= A$, which means for any $x\in A$, $(2x\pmod n)\in A$. We can conclude that $C_x^{(2,n)}\subseteq A$. Then $A$ is a union of some cyclotomic sets of $\mathbb{Z}_n$.
      \end{IEEEproof}
 According to the definition of the defining set mentioned above, we have the following Corollary.
 \begin{corollary}\label{cor1}
     Let $n=2^m-1$. Suppose $Z\subseteq \mathbb{Z}_n$. Then $Z$ is a defining set of a binary cyclic code with length $n$ if and only if $\rho(\pi(Z))=\pi(Z).$
 \end{corollary}

\section{First Construction: \texorpdfstring{$m$}{m} is even}\label{set00}
In this section, we will consider the construction of binary cyclic codes of length $n=2^m-1$ when $m$ is even.
In \cite{ref4}, Tang et al. constructed a defining set $A$ such that $w_2(i)$ are either all odd or all even for every $i\in A$, where $w_2(i)=\sum_{j=0}^{m-1}i_j$ and $i=\sum_{j=0}^{m-1}i_j2^j$. In this paper, we propose another idea to obtain the defining set. Our method is inspired by the construction of rotation symmetric Boolean functions in \cite{ref2}.

Let $m=2t$. For every $\bm{x}=(x_0,x_1,...,x_{2t-1})\in\mathbb{F}_2^m$, we define $\bm{x}'=(x_0,x_1,...,x_{t-1})$ and $\bm{x}''=(x_t,x_{t+1},...,x_{2t-1}).$ Then $\bm{x}=(\bm{x}',\bm{x}'')$. Let ${\bm{1}_a}=(1,1,...,1)\in\mathbb{F}_2^a$ and ${\bm{0}_a}=(0,0,...,0)\in\mathbb{F}_2^a$. Further, we denote $W^{(m)}_{>t}=\left\{\bm{x}\in\mathbb{F}_2^m|wt(\bm{x})>t\right\}$, $W^{(m)}_{t}=\left\{\bm{x}\in\mathbb{F}_2^m|wt(\bm{x})=t\right\}$ and $W^{(m)}_{<t}=\left\{\bm{x}\in\mathbb{F}_2^m|wt(\bm{x})<t\right\}.$ Then we can get the following sets:
\begin{align*}
    B&=\left\{\bm{x}\in W^{(m)}_{>t} \mid \bm{x}'=\bm{x}''\right\}=\left\{\bm{x}\in W^{(m)}_{>t} \mid \rho^t(\bm{x})=\bm{x}\right\},\\
    T&=\left\{\bm{x}\in W^{(m)}_{t} \mid \bm{x}'=\bm{x}''\right\}=\left\{\bm{x}\in W^{(m)}_{t} \mid \rho^t(\bm{x})=\bm{x}\right\},\\
    G&=\left\{\bm{x}\in W^{(m)}_{t} \mid \bm{x}'=\bm{x}''+\bm{1}_t\right\}=\left\{\bm{x}\in W^{(m)}_{t} \mid \rho^t(\bm{x})=\bm{x}+\bm{1}_m\right\}.
\end{align*}

The following lemmas provide some good properties of the sets $B$, $T$ and $G$, which are obtained in \cite{ref2}. For completeness, we present their proofs here.
\begin{lemma}[\hspace{-0.01em}\cite{ref2}]\label{lem2}
    If $\bm{x}\in B$ (respectively $T$ and $G$), then $\rho(\bm{x})\in B$ (respectively $T$ and $G$), where the subsets $B$, $T$ and $G$ are defined above.
\end{lemma}
\begin{IEEEproof}
If $\bm{x}=(x_0,x_1,...,x_{t-1},x_0,x_1,...,x_{t-1})\in B$, we have
    $$\rho(\bm{x})=(x_{t-1},x_0,x_1,...,x_{t-2},x_{t-1},x_0,...,x_{t-2}),$$
    which leads to $(\rho(\bm{x}))'=(\rho(\bm{x}))''$. Then $\rho(\bm{x})\in B$ since $wt({\bm{x}})=wt(\rho({\bm{x}}))$.
    
    Similarly, we can obtain $\rho(\bm{x})\in T$ for any $x\in T$.
    
 For any $\bm{x}=(x_0,x_1,...,x_{t-1},x_0+1,x_1+1,...,x_{t-1}+1)\in G$, we have 
 \begin{align*}
     \rho({\bm{x}})=&(x_{t-1}+1,x_0,x_1,...,x_{t-2},x_{t-1},x_0+1,x_1+1,...,x_{t-2}+1).
 \end{align*}
   
we can get $(\rho({\bm{x}}))'=(x_{t-1}+1,x_0,x_1,...,x_{t-2})$ and $(\rho({\bm{x}}))''=(x_{t-1},x_0+1,x_1+1,...,x_{t-2}+1),$ which means $(\rho({\bm{x}}))'=(\rho({\bm{x}}))''+\bm{1}_t.$ It leads to $\rho(\bm{x})\in G$.
\end{IEEEproof}

Due to $wt({\bm{x}})=wt(\rho({\bm{x}}))$, we can conclude that if $\bm{x}\in W_{t}^{(m)}$(respectively $W_{>t}^{(m)}$, $W_{<t}^{(m)}$), then $\rho(\bm{x})\in W_{t}^{(m)}$ (respectively $W_{>t}^{(m)}$, $W_{<t}^{(m)}$). As for the subset $W_{t}^{(m)}\backslash(T\cup G)$, we have the following lemmas.
\begin{lemma}[\hspace{-0.01em}\cite{ref2}]\label{lem3}
    Let the notation remain as above. For any $\bm{x}\in W_t^{(m)}\backslash(T\cup G)$ and $0\le l\le m-1$, we have $\rho(\bm{x}+\bm{1}_m)\in W_t^{(m)}\backslash(T\cup G)$. 
\end{lemma}
\begin{IEEEproof}
    Clearly, if $\bm{x}\in W_t^{(m)}$, then $wt(\bm{x})=t$. It leads to $wt(\bm{x}+\bm{1}_m)=m-t=t$, Thus, we have $\bm{x}+\bm{1}_m\in W_t^{(m)}$ and $\rho({\bm{x}+\bm{1}}_m)\in W_t^{(m)}.$

    Further, for any $\bm{x}\not\in T\cup G$, assume that $\bm{x}+\bm{1}_m\in T\cup G$. By the definition of $T$ and $G$, we then have $(\bm{x}+\bm{1}_m)+\bm{1}_m\in T\cup G$, which implies that $\bm{x}\in T\cup G$, contradicting the assumption. Thus, by Lemma \ref{lem2}, we have $\rho({\bm{x}+\bm{1}}_m)\not\in T\cup G.$ Therefore, the conclusion follows.
\end{IEEEproof}

\begin{lemma}[\hspace{-0.01em}\cite{ref2}]\label{lem6}
    For any $0\le l\le m-1$, there is no $\bm{x}\in W_t^{(m)}\backslash(T\cup G)$ such that $\bm{x}+\bm{1}_m=\rho^l(\bm{x}).$
\end{lemma}

\begin{IEEEproof}
    We suppose that $l_0$ is the minimal positive integer such that there exists $x\in W_t^{(m)}\backslash(T\cup G)$ satisfying ${\bm{x}}+{\bm{1}}_m=\rho^{l_0}(\bm{x}).$ Then
\begin{align*}
    \bm{x}&=\bm{x}+\bm{1}_m+\bm{1}_m=\rho^{l_0}(\bm{x})+\bm{1}_m=\rho^{l_0}(\bm{x}+\bm{1}_m)
    =\rho^{l_0}(\rho^{l_0}(\bm{x}))=\rho^{2l_0}(\bm{x}).
\end{align*}
    
We claim that $2l_0\mid m.$ In fact, let $m=2l_0u+v$ with $0 \leq v<2l_0$, then we have $\rho^v({\bm{x}})=\bm{x}.$ If $0 <v<l_0$ then $\rho^{l_0-v}({\bm{x}})=\bm{x}+\bm{1}_m.$ If $l_0 \leq v<2l_0$, then $\rho^{2l_0-v}({\bm{x}})=\bm{x}$ and $\rho^{v-l_0}({\bm{x}})=\bm{x}+\bm{1}_m,$ which contradicts to the minimality of $l_0$. Thus, $v=0$ and $2l_0 \mid m$. Therefore, we conclude that  $$\bm{x}=(\bm{\alpha},\bm{\alpha}+\bm{1}_{l_0},\bm{\alpha},\bm{\alpha}+\bm{1}_{l_0},...,\bm{\alpha}+\bm{1}_{l_0})$$
        for some $\bm{\alpha}\in\mathbb{F}_2^{l_0}$. 
    If $\frac{m}{2l_0}$ is even, then $\bm x'=\bm x''$, implying that $\bm{x}\in T$. If $\frac{m}{2l_0}$ is odd, then $\bm{x}'=\bm{x}''+\bm{1}_t,$ which implies that $\bm{x}\in G.$ Thus, we conclude that $\bm{x}\in T\cup G$, leading to a contradiction. The lemma is proved.
\end{IEEEproof}
For a subset $A\subseteq \mathbb{F}_2^m$, define $A+\bm{1}_m=\left\{\bm{x}+\bm{1}_m:\bm{x}\in A\right\}$ and $\rho(A)=\{\rho(\bm x): \bm x \in A\}$. According to Lemmas \ref{lem3} and \ref{lem6}, we can conclude that there exist $P_1,P_2\subseteq W_t^{(m)}\backslash(T\cup G)$ such that 
    $$\begin{cases}
        &\rho(P_i)= P_i\ \ (i=1,2),\\
        &P_1=P_2+\bm{1}_m,\\
        &P_1\bigcap P_2=\emptyset,\\
        &P_1\bigcup P_2=W_t^{(m)}\backslash(T\cup G).
    \end{cases}$$
Let 
 $$\begin{cases}
     &S_1=W_{<t}^{(m)}\bigcup B\bigcup T\bigcup P_1,\\
     &S_2=(W_{>t}^{(m)}\backslash B)\bigcup G\bigcup P_2.
 \end{cases}$$
 According to the definition above, we obtain that $S_1\cap S_2=\emptyset$, $S_1\cup S_2=\mathbb{F}_2^m$ and $\pi(0)=\bm{0}_m,\pi(n)=\bm{1}_m\in S_1,$ where $\pi$ is defined in Section \ref{set2}.
 Let
 \begin{equation}\label{eq0}
     Z_1=\pi^{-1}(S_1)\backslash\left\{0,n\right\},
 \end{equation}
 and
 \begin{equation}\label{eq00}
     Z_2=\pi^{-1}(S_2).
 \end{equation}
Then $Z_1, Z_2 \subseteq \mathbb{Z}_n$. By Lemmas \ref{lem2}, \ref{lem3} and \ref{lem6}, we have $\rho(\bm{x})\in S_i$ for any $\bm{x}\in S_i$ $(i=1,2)$. Thus, by Corollary \ref{cor1}, both sets 
$Z_1$ and $Z_2$ are defining sets of binary cyclic codes. Consequently, we obtain the following theorem.
\begin{theorem}\label{thm4}
    Let $n=2^m-1$ with even $m$. The binary cyclic code $C_1$, defined by the defining set $Z_1$ in (\ref{eq0}),
    has parameters \[[2^m-1,2^{m-1}+1,d\ge 2^{m/2}-1].\]
    Its dual code $C_1^\perp$ has parameters 
    \[[2^m-1,2^{m-1}-2,d^\perp\ge 2^{m/2}].\] 
    Similarly, the binary cyclic code $C_2$, defined by the defining set $Z_2$ in (\ref{eq00}),
    has parameters 
    \[[2^m-1,2^{m-1}-1,d\ge 2^{m/2}-1],\] and its dual code $C_2^\perp$ has  parameters 
    \[[2^m-1,2^{m-1},d^\perp\ge 2^{m/2}].\] 
\end{theorem}
\begin{IEEEproof}
Clearly, $|S_1|=|W_{<t}^{(m)}|+|B|+|T|+|P_1|$ and $|S_2|=|W_{>t}^{(m)}|+|G|+|P_2|-|B|$ by the definition above.
  
By the definitions of the subsets $B$, $T$ and $G$,  we can derive that
 $$
 |B|=\sum_{\lceil \frac{t+1}{2}\rceil}^t\binom{t}{i}=\left\{\begin{aligned}
    &2^{t-1},\ \ t\text{ is odd},\\
    &2^{t-1}-\frac{1}{2}\binom{t}{\frac{t}{2}},\ \ t\text{ is even}.
 \end{aligned}
    \right.$$

    $$|T|=\left\{\begin{aligned}
        &0,\ \ t\text{ is odd},\\
        &\binom{t}{\frac{t}{2}},\ \ t\text{ is even},
    \end{aligned}
    \right.$$
and $|G|=2^t$ . Since $P_1=P_2+\bm{1}_m$, then $|P_1|=|P_2|$, we have $|P_1|=|P_2|=\frac{1}{2}(|W_{t}^{(m)}|-|G|-|T|)$, which leads to
    $$|P_1|=|P_2|=\left\{\begin{aligned}
        &\frac{1}{2}\left(\binom{m}{t}-2^t\right),\ \ t\text{ is odd},\\
        &\frac{1}{2}\left(\binom{m}{t}-\binom{t}{\frac{t}{2}}-2^t\right),\ \ t\text{ is even}.
    \end{aligned}
       \right.$$
   Then we have $|B|+|T|+|P_1|=\frac{1}{2}\binom{m}{t}$ and $|G|+|P_2|-|B|=\frac{1}{2}\binom{m}{t}.$

Note that $|W_{<t}^{(m)}|=|W_{>t}^{(m)}|$ since $W_{<t}^{(m)}+\bm{1}_m=W_{>t}^{(m)}$. Thus, 
\begin{align*}
    |S_1|=&|W_{<t}^{(m)}|+|B|+|T|+|P_1|=|W_{>t}^{(m)}|+|G|+|P_2|-|B|=|S_2|.
\end{align*}

Then we can conclude that $|Z_1|=2^{m-1}-2$ and $|Z_2|=2^{m-1}$, which leads to ${\rm{dim}}(C_1)=2^{m-1}+1$ and ${\rm{dim}}(C_2)=2^{m-1}-1$.

 Since $wt(\pi(i))<t$ for any $1\le i\le 2^t-2$, we have 
\begin{align*}
    \left\{1,2,...,2^t-2\right\}&\subseteq\pi^{-1}(W_{<t}^{(m)})\backslash\left\{0\right\}\subseteq\pi^{-1}(S_1)\backslash\left\{0,n\right\}=Z_1.
\end{align*}
By the BCH bound, we get $d(C_1)\ge 2^t-1$.
   
On the other hand, for any $i$ with $2^m-2^t+1\le i\le2^m-2$, we have $\pi(i)=({\bm{a}},\bm{1}_t)$ with ${\bm{a}}\in \mathbb F_{2}^t\backslash\left\{\bm{0}_t,\bm{1}_t\right\}.$ Thus, $wt(\pi(i))>t$ and $\pi(i)\cap B=\emptyset$, which leads to $\pi(i)\in W_{>t}^{(m)}\backslash B.$ Then we can conclude that $i\in Z_2$, therefore
$$\left\{2^m-2^t+1,2^m-2^t+2,...,2^m-2\right\}\subseteq Z_2.$$
 From the BCH bound, $d(C_2)\ge 2^t-1.$

  Note that $Z_1\cup Z_2=\mathbb{Z}_n\backslash\left\{0\right\}$, then $-(\mathbb{Z}_n\backslash Z_1)=\left\{0\right\}\cup (-Z_2)$ and $-(\mathbb{Z}_n\backslash Z_2)=\left\{0\right\}\cup (-Z_1)$. Thus, we have 
  $$\left\{0,1,2...,2^t-2\right\}\subseteq-(\mathbb{Z}_n\backslash Z_1)$$ and 
  \begin{align*}
      \left\{2^m-2^t+1,2^m-2^t+2,...,2^m-2,2^m-1\right\}\subseteq-(\mathbb{Z}_n\backslash Z_2).
  \end{align*}
   From the BCH bound, we know that the minimum distance $d(C_1^\perp)\ge 2^t$ and $d(C_2^\perp)\ge 2^t.$
   
   The proof is completed.
\end{IEEEproof}

\begin{remark}
    As far as we know, few studies have focused on binary cyclic codes of length $n=2^m-1$, dimension near $n/2$, and large minimum distance when $m$ is even. In \cite{ref4}, Tang et al. constructed some binary cyclic codes with parameters $[2^m-1,2^{m-1}\pm1,d]$ for even $m$. We list the parameters of these constructions in the following Table \ref{table3}. It can be seen that our binary cyclic codes have better minimum distance and dual distance compared to theirs.
\end{remark}
\begin{table*}[ht]
            \renewcommand{\arraystretch}{1.5}
            \centering
            \caption{Comparison between the lower bounds on minimum distance in \cite{ref4} and in Theorem \ref{thm4} Binary $[n,k]$ cyclic codes with minimum distance $d$ and dual distance $d^{\perp}$ when $m$ is even\centering}
            \label{table3}
            \begin{tabular}{|c|c|c|c|c|}
             \hline
                $m$&$k$& $d\ge$  & $ d^\perp\ge$  & Reference\\
                \hline
                $m\equiv4\pmod 8$&$2^{m-1}+1$&$2^{(m-4)/2}+1$&$2^{(m-4)/2}+2$&\cite{ref4}\\
                \hline
                $m\equiv2\pmod 4$&$2^{m-1}+1$&$2^{(m-2)/2}+1$&$2^{(m-2)/2}+2$&\cite{ref4}\\
                \hline
                $m\equiv0\pmod 4$&$2^{m-1}-1$&$2^{(m-2)/2}+1$&$2^{(m-2)/2}+2$&\cite{ref4}\\
                \hline
                $m\equiv2\pmod 4$&$2^{m-1}-1$&$2^{(m-4)/2}+1$&$2^{(m-4)/2}+2$&\cite{ref4}\\
                \hline
                $m$ is even&$2^{m-1}-1$&$2^{\frac{m}{2}}-1$& $2^{\frac{m}{2}}$&Theorem \ref{thm4} \\
                \hline
             $m$ is even&$2^{m-1}+1$&$2^{\frac{m}{2}}-1$& $2^{\frac{m}{2}}$&Theorem \ref{thm4} \\
             \hline
 \end{tabular}
        \end{table*}

\begin{example}
    Let $m=4$ and $n=2^m-1$ then $t=2$. We can compute that 
    \begin{align*}
        &\pi^{-1}(W_{<t}^{(m)})=\left\{0,1,2,4,8\right\},\\
        &\pi^{-1}(W_{t}^{(m)})=\left\{3,5,6,9,10,12\right\},\\
        &\pi^{-1}(W_{>t}^{(m)})=\left\{7,11,13,14,15\right\},\\
        &\pi^{-1}(B)=\left\{15\right\},\\
        &\pi^{-1}(T)=\left\{5,10\right\},\\
        &\pi^{-1}(G)=\left\{3,6,9,12\right\}.
        \end{align*}
    In this case, $P_1=P_2=\emptyset$, leading to the defining sets 
 $Z_1=\left\{1,2,4,5,8,10\right\}$ and $Z_2=\left\{3,6,9,11,12,13,14\right\}$.
    Then the binary cyclic code $C_1$ with defining set $Z_1$ has parameters $[15,9,3]$, while its dual $C_1^{\perp}$ has parameters $[15,6,6]$ and attains the best known parameters\cite{ref41}. Similarly, $C_2$ with defining set $Z_2$ has parameters $[15,7,5]$ and its dual $C_1^{\perp}$ has parameters $[15,8,4]$, with both $C_2$ and $C_2^\perp$ achieving the best known parameters.
\end{example}

\begin{example}
Let $m=6$ and $n=2^m-1$ then $t=3$. We can compute that
\begin{align*}
    \pi^{-1}(W_{<t}^{(m)})=&\left\{0,1,2,3,4,5,6,8,9,10,12,16,17,18,20,\right.\left.24,32,33,34,36,40,48\right\},\\
    \pi^{-1}(W_{t}^{(m)})=&\left\{7,11,13,14,19,21,22,25,26,28,35,37,\right.\left.38,41,42,44,49,50,52,56\right\},\\
    \pi^{-1}(W_{>t}^{(m)})=&\left\{15,23,27,29,30,31,39,43,45,46,47,51,\right.\left.53,54,55,57,58,59,60,61,62,63\right\},\\
    \pi^{-1}(B)=&\left\{27,45,54,63\right\},\\
    \pi^{-1}(T)=&\emptyset,\\
    \pi^{-1}(G)=&\left\{7,14,21,28,35,42,49,56\right\}.
\end{align*}

In this case, $W_t^{(m)}\backslash G=\left\{11,13,19,22,25,26,37,38,41,\right.\left.44,50,52\right\}$, then we can get $\pi^{-1}(P_1)=\left\{13,19,26,38,41,52\right\}$ and $\pi^{-1}(P_2)=\left\{11,22,25,37,44,50\right\}.$ It leads to 
\begin{align*}
    Z_1=&\left\{1,2,3,4,5,6,8,9,10,12,13,16,17,18,19,20,24,26,27,32,33,34,36,38,40,41,45,48,52,54\right\}
\end{align*}

and
\begin{align*}
    Z_2&=\left\{11,14,15,21,22,23,25,28,29,30,31,35,37,39,42,43,44,46,47,49,50,51,53,55,56,57,58,59,60,61,62\right\}.
\end{align*}

 Then the binary cyclic code $C_1$ with defining set $Z_1$ has parameters $[63,33,7]$ and its dual $C_1^{\perp}$ has parameters $[63,30,12]$. Similarly, $C_2$ with defining set $Z_2$ has parameters $[63,31,9]$ and its dual $C_2^{\perp}$ has parameters $[63,32,8].$
\end{example}

\section{Second Construction: \texorpdfstring{$m=p_1p_2$}{m=p1p2}}\label{sect4}
In \cite{ref7}, Sun et al. presented a family of binary cyclic codes with length $n=2^m-1$ when $m$ is a prime. Further, its minimum distance can exceed $\frac{n}{\log_2n}$. In this section, we consider the construction of binary cyclic codes of length $n=2^m-1$ for $m=p_1p_2$ where $p_1$ and $p_2$ are different primes. To begin with, we present some results about the cyclotomic coset $C_x^{(2,n)}.$
\begin{lemma}\label{lem5}
Let $n=2^m-1$. For any $0<x\le n-1$ and $\pi(x)=(x_0,x_1,...,x_{m-1})$, then $x\equiv x\cdot2^{t}\pmod n$ if and only if $x_j=x_{j+t\pmod m}$ for any $0\le j\le m-1$. Moreover, suppose $t$ is a prime and $t\mid m,$ then $\left|C_x^{(2,n)}\right|=t$ if and only if 
 $$x=a\sum_{i=0}^{\frac{m}{t}-1}2^{it}$$ for some $1\le a \le 2^{t}-2.$ 
\end{lemma}
\begin{IEEEproof}
    By the proof of the Lemma \ref{lem1}, we know that 

    \begin{align*}
        &\pi(x\cdot 2^{t}\pmod n)=\rho^{t}(x_0,x_1,...,x_{m-1})\\=&(x_{m-t},x_{m-t+1},...,x_{m-1},x_0,x_1,...,x_{m-t-1}),
    \end{align*}
  where the subscripts are indexed in modulo $m$. The first conclusion follows.

If $\left|C_x^{(2,n)}\right|=t$, then $x\equiv x\cdot2^t\pmod{n}$. By the first conclusion we know that $\pi(x)=(\underbrace{{\bm{b}},{\bm{b}},...,{\bm{b}}}_\frac{m}{t})$ for some ${\bm{b}}\in\mathbb{F}_2^t\backslash\left\{\bm{0}_t,\bm{1}_t\right\}.$ Thus, we can conclude that
$$x=a\sum_{i=0}^{\frac{m}{t}-1}2^{it},$$ where $\pi(a)=({\bm{b}},\underbrace{0,0,..,0}_{m-t}),$ which means $1\le a\le 2^{t}-2.$

Conversely, if $x=a\sum_{i=0}^{\frac{m}{t}-1}2^{it}$ for some $1\le a\le 2^t-2$,
then $\pi(a)=(a_0,a_1,...,a_{t-1},\underbrace{0,0,...,0}_{m-t}),$ where $a_i\in\mathbb{F}_2(i=0,1,...,t-1)$. It leads to $\pi(x)=(a_0,a_1,...,a_{t-1},a_0,a_1,...,a_{t-1},...,a_0,a_1,...,a_{t-1}),$ which means $x_j=x_{j+t\pmod m}.$
  Suppose $\left|C_x^{(2,n)}\right|=t'<t$, then $x_j=x_{j+t'\pmod m}.$ Since $t$ is a prime, we can conclude that $x_j=x_{j+1\pmod m}$. It leads to $x=n$, thus, $t'\ge t$. On the other hand, $x\equiv x\cdot 2^t\pmod n$ leads to $t\ge t'$. Therefore, $t=t'$.
\end{IEEEproof}

It is well known that if $m=p_1p_2$ and $0<x\le n-1$, the cardinality of the set $C_x^{(2,n)}$ can only be $p_1$, $p_2$ or $p_1p_2$. By Lemma \ref{lem5}, we obtain the characterization of $C_x^{(2,n)}$ according to its size in the following Corollary.
\begin{corollary}\label{cor3}
    Let $m=p_1p_2$ and $n=2^m-1$ where $p_1$ and $p_2$ are two different primes. Suppose $x\in\mathbb{Z}_n\backslash\left\{0\right\}$, then 
    \begin{itemize}
     \item[1)] $\left|C_x^{(2,n)}\right|=p_1$ if and only if $x=a\sum_{i=0}^{p_2-1}2^{ip_1}$ for some $1\le a\le 2^{p_1}-2$ if and only if $\pi(x)=(\underbrace{{\bm{b}},{\bm{b}},...,{\bm{b}}}_{p_2})$ for some ${\bm{b}}\in\mathbb{F}_2^{p_1}\backslash\left\{\bm{0}_{p_1},\bm{1}_{p_1}\right\}.$ Furthermore, 
     
     $$\left|\left\{C_x^{(2,n)}:\left|C_x^{(2,n)}\right|=p_1\right\}\right|=\frac{2^{p_1}-2}{p_1},$$
     \item[2)] $\left|C_x^{(2,n)}\right|=p_2$ if and only if $x=a\sum_{i=0}^{p_1-1}2^{ip_2}$ for some $1\le a\le2^{p_2}-2$ if and only if $\pi(x)=(\underbrace{{\bm{b}},{\bm{b}},...,{\bm{b}}}_{p_1})$ for some ${\bm{b}}\in\mathbb{F}_2^{p_2}\backslash\left\{\bm{0}_{p_2},\bm{1}_{p_2}\right\}.$ Furthermore, 
     $$\left|\left\{C_x^{(2,n)}:\left|C_x^{(2,n)}\right|=p_2\right\}\right|=\frac{2^{p_2}-2}{p_2},
     $$
     \item[3)] $\left|C_x^{(2,n)}\right|=p_1p_2$ if and only if both $\sum_{i=0}^{p_2-1}2^{ip_1}\nmid x$ and $\sum_{i=0}^{p_1-1}2^{ip_2}\nmid x$. Furthermore, 
     
     $$\left|\left\{C_x^{(2,n)}:\left|C_x^{(2,n)}\right|=p_1p_2\right\}\right|=\frac{2^{p_1p_2}-2^{p_1}-2^{p_2}+2}{p_1p_2}.$$
\end{itemize}
\end{corollary}
Note that when $p_1=2$, $\left|\left\{C_x^{(2,n)}:\left|C_x^{(2,n)}\right|=p_1\right\}\right|=\frac{2^{p_1}-2}{p_1}=1$. When $p_1$ is an odd prime, $\left|\left\{C_x^{(2,n)}:\left|C_x^{(2,n)}\right|=p_1\right\}\right|=\frac{2^{p_1}-2}{p_1}$ is even. Therefore, we divide our discussion into two cases: $m=2p$ and $m=p_1p_2$ $(p_2>p_1\ge3).$

\subsection{\texorpdfstring{$m=2p$}{m=2p}}\label{set1}
In this section, we suppose $m=2p$ and $n=2^m-1=2^{2p}-1$, where $p$ is an odd prime.

By Corollary \ref{cor3}, we can obtain that $\left|C_x^{(2,n)}\right|=2$ if and only if $x=\sum_{i=0}^{p-1}2^{2i}=\frac{1}{3}(2^{2p}-1)$ or $x=\sum_{i=0}^{p-1}2^{2i+1}=\frac{2}{3}(2^{2p}-1)$. Moreover,  \[\left|\left\{C_x^{(2,n)}:\left|C_x^{(2,n)}\right|=p\right\}\right|=\frac{2^{p}-2}{p}\]
and 
\[\left|\left\{C_x^{(2,n)}:\left|C_x^{(2,n)}\right|=2p\right\}\right|=\frac{2^{2p}-2^{p}-2}{2p}.\]
Let $1\le i_1<i_2<\cdots<i_{\frac{2^p-2}{p}}\le n-1$ be the coset leaders of the 2-cyclotomic cosets $C_{i_\alpha}^{(2,n)}$ with $\left|C_{i_\alpha}^{(2,n)}\right|=p$ $(1\le\alpha\le\frac{2^p-2}{p})$ and $1\le j_1<j_2<\cdots<j_{\frac{2^{2p}-2^p-2}{2p}}\le n-1$ be the coset leaders of the 2-cyclotomic cosets $C_{j_\beta}^{(2,n)}$ with $\left|C_{j_\beta}^{(2,n)}\right|=2p$ $(1\le \beta\le\frac{2^{2p}-2^p-2}{2p})$.

Let $\sigma$ be an integer with $\frac{2^{2p}-2^{p+1}}{4p}\le\sigma\le\frac{2^{2p}-4}{4p}<\frac{2^{2p}-2^p-2}{2p}$, then
    $0 \leq \frac{2^{2p}-4}{2p}-2\sigma\le\frac{2^p-2}{p}$.
For any $1\le\beta_1<\beta_2<\cdots<\beta_\sigma\le\frac{2^{2p}-2^p-2}{2p}$, we define
$$Z_1(\sigma;\beta_1,...,\beta_\sigma)=\left(\bigcup_{l=1}^{\sigma}C^{(2,n)}_{j_{\beta_l}}\right)\bigcup\left(\bigcup_{\alpha=1}^{\frac{2^{2p}-4}{2p}-2\sigma}C_{i_\alpha}^{(2,n)}\right).$$
Then
\begin{align*}
    \left|Z_1(\sigma;\beta_1,...,\beta_\sigma)\right|&=2p\sigma+p\cdot\left(\frac{2^{2p}-4}{2p}-2\sigma\right)\\&=\frac{2^{2p}-4}{2}=2^{m-1}-2.
\end{align*}

\begin{theorem}\label{thm1}
Let $p$ be an odd prime and $m=2p$, then there exists a $[2^m-1,2^{m-1},d]$ binary cyclic code $C$ with $d\ge \frac{2^{2p}-4}{2p}+\frac{2^p-2p-2}{4p}+1$.

\end{theorem}
\begin{IEEEproof}
    Note that each coset leader of a 2-cyclotomic coset is odd. For any subset $D\subseteq\mathbb{Z}_n$, we define 
    
    $$D_{odd}=\left\{i:i\in D, i\equiv1\pmod2\right\},$$
then $$\bigcup_{i\in D}C_i^{(2,n)}=\bigcup_{i\in D_{odd}}C^{(2,n)}_i.$$
    Consider the set 
  \begin{align*}
      D_0=&\left\{1,2,...,\frac{2^{2p}-4}{2p},\frac{2^{2p}-4}{2p}+1,...,\frac{2^{2p}-4}{2p}+\frac{2^p-2p-2}{4p}-1\right\}.
  \end{align*}  
        If $p=3$, then $\frac{2^{p}-2}{2p}(2^p+1)=\frac{2^{2p}-4}{2p}+\frac{2^p-2p-2}{4p}-1=9$.
If $p\ge5$, then $\frac{2^{p}-2}{2p}(2^p+1)=\frac{2^{2p}-2^p-2}{2p}<\frac{2^{2p}-4}{2p}\le\frac{2^{2p}-4}{2p}+\frac{2^p-2p-2}{4p}-1$.
    Thus, we have 
    \begin{align*}
        D_1\triangleq\left\{a(2^p+1):a=1,2,...,\frac{2^{p}-2}{2p}\right\}\subseteq \left\{x:x\in D_0,\left|C_x^{(2,n)}\right|=p\right\}\subseteq D_0.
    \end{align*}
  
Since $\frac{2^{p}-2}{2p}$ is odd, then
$$ |(D_1)_{odd}|=
\frac{1}{2}\left(\frac{2^{p}-2}{2p}+1\right),$$
and
\begin{align*}
    \left|(D_0\backslash D_1)_{odd}\right|=\left\lfloor\frac{1}{2}\left(\frac{2^{2p}-4}{2p}+\frac{2^p-2p-2}{4p}\right)\right\rfloor -\frac{1}{2}\left(\frac{2^{p}-2}{2p}+1\right).
\end{align*}

Note that $\frac{1}{3}(2^{2p}-1)>\frac{2^{2p}-4}{2p}+\frac{2^p-2p-2}{4p}-1$. Then, by Corollary \ref{cor3}, for any $x\in D_0$, the value of $|C_x^{(2,n)}|$ can only be $2p$ or $p$. Since for any $x\in D_1\subseteq D_0$, $|C_x^{(2,n)}|=p$, then   
\begin{align*}
    \left|\bigcup_{i\in D_0}C_i^{(2,n)}\right|\le& 2p\cdot\left|(D_0\backslash D_1)_{odd}\right|+p\cdot\left|(D_1)_{odd}\right|\\
    \le&2p\cdot\left[\frac{1}{2}\left(\frac{2^{2p}-4}{2p}+\frac{2^p-2p-2}{4p}\right)-\frac{1}{2}\left(\frac{2^p-2}{2p}+1\right)\right]+p\cdot\frac{1}{2}\left(\frac{2^p-2}{2p}+1\right)\\
    =& \frac{2^{2p}-4}{2}-p\cdot\frac{1}{2}\left(\frac{2^p-2}{2p}+1\right)
    +\frac{2^p-2p-2}{4}\\
    =&\frac{2^{2p}-4}{2}-p<\frac{2^{2p}-4}{2}.
\end{align*}

Let $\sigma_0$ be the smallest integer such that

 $$\left\{x:x\in D_0, |C_x^{(2,n)}|=2p\right\}\subseteq\bigcup_{\beta=1}^{\sigma_0}C_{j_\beta}^{(2,n)}.$$

Then $\left\{x:x\in \bigcup_{x\in D_0}C_{x}^{(2,n)}, \left|C_x^{(2,n)}\right|=2p\right\}=\bigcup_{\beta=1}^{\sigma_0}C_{j_\beta}^{(2,n)}.$
By the above, $\sigma_0\le\frac{2^{2p}-4}{4p}$. 

If $\frac{2^{2p}-2^{p+1}}{4p}\le\sigma_0\le\frac{2^{2p}-4}{4p}$,  then

    \begin{equation}\label{6}
        \begin{aligned}
             \bigcup_{x\in D_0}C_{x}^{(2,n)}=&\left\{x:x\in \bigcup_{x\in D_0}C_{x}^{(2,n)}, \left|C_x^{(2,n)}\right|=2p\right\}\bigcup\left\{x:x\in \bigcup_{x\in D_0}C_{x}^{(2,n)}, \left|C_x^{(2,n)}\right|=p\right\}\\
             =&\left\{x:x\in \bigcup_{x\in D_0}C_{x}^{(2,n)}, \left|C_x^{(2,n)}\right|=p\right\}
            \bigcup\left(\bigcup_{l=1}^{\sigma_0}C_{j_{l}}^{(2,n)}\right)\\
    \stackrel{\rm (a)}{\subseteq}&\left(\bigcup_{l=1}^{\sigma_0}C_{j_{l}}^{(2,n)}\right)\bigcup\left(\bigcup_{\alpha=1}^{\frac{2^{2p}-4}{2p}-2\sigma_0}C_{i_\alpha}^{(2,n)}\right)\\
    =&Z_1(\sigma_0;1,2,...,\sigma_0),
        \end{aligned}
    \end{equation}
   where $\rm (a)$ follows from $\left|\bigcup_{x\in D_0}C_{x}^{(2,n)}\right|\le\frac{2^{2p}-4}{2}=
   \left|\left(\bigcup_{l=1}^{\sigma_0}C_{j_{\beta_l}}^{(2,n)}\right)\bigcup\left(\bigcup_{\alpha=1}^{\frac{2^{2p}-4}{2p}-2\sigma_0}C_{i_\alpha}^{(2,n)}\right)\right|.$

If $\sigma_0<\frac{2^{2p}-2^{p+1}}{4p}$, then for any  $\sigma_0 <\beta_{\sigma_0+1}<\beta_{\sigma_0+2}<\cdots<\beta_{\frac{2^{2p}-2^{p+1}}{4p}}\le\frac{2^{2p}-2^p-2}{2p}$, we have
\begin{align*}
    \bigcup_{x\in D_0}C_{x}^{(2,n)}=&\left\{x:x\in \bigcup_{x\in D_0}C_{x}^{(2,n)}, \left|C_x^{(2,n)}\right|=2p\right\}\bigcup\left\{x:x\in \bigcup_{x\in D_0}C_{x}^{(2,n)}, \left|C_x^{(2,n)}\right|=p\right\}\\
    \subseteq&\left(\bigcup_{l=1}^{\sigma_0}C_{j_{l}}^{(2,n)}\right)\bigcup\left(\bigcup_{\alpha=1}^{\frac{2^p-2}{p}}C_{i_\alpha}^{(2,n)}\right)\\
    \subseteq& \left(\bigcup_{l=1}^{\sigma_0}C_{j_{l}}^{(2,n)}\right)\bigcup\left(\bigcup_{l=\sigma_0+1}^{\frac{2^{2p}-2^{p+1}}{4p}}C_{j_{\beta_l}}^{(2,n)}\right) \bigcup\left(\bigcup_{\alpha=1}^{\frac{2^p-2}{p}}C_{i_\alpha}^{(2,n)}\right)\\
    =&Z_1\left(\frac{2^{2p}-2^{p+1}}{4p};1,2,...,\sigma_0,\beta_{\sigma_0+1},...,\beta_{\frac{2^{2p}-2^{p+1}}{4p}}\right).
\end{align*}

Thus, we can always choose an integer $\sigma$ with $\frac{2^{2p}-2^{p+1}}{4p}\le\sigma\le\frac{2^{2p}-4}{4p}$ and $1\leq \beta_1 <\beta_2<...<\beta_\sigma \leq \frac{2^{2p}-2^p-2}{2p}$ such that
\begin{align*}
    D_0&=\left\{1,2,...,\frac{2^{2p}-4}{2p},\frac{2^{2p}-4}{2p}+1,...,\frac{2^{2p}-4}{2p}+\frac{2^p-2p-2}{4p}-1\right\}\subseteq Z_1(\sigma;\beta_1,...,\beta_\sigma).
\end{align*}

Let $C$ be the binary cyclic code with defining set $Z_1(\sigma;\beta_1,...,\beta_\sigma)\cup\left\{0\right\}$. Then ${\rm{dim}}(C)=n-|Z_1(\sigma;\beta_1,...,\beta_\sigma)|-1=2^{m-1}$. Since $\left\{0,1,2,...,\frac{2^{2p}-4}{2p},\frac{2^{2p}-4}{2p}+1,...,\frac{2^{2p}-4}{2p}+\frac{2^p-2p-2}{4p}-1\right\}\subseteq Z_1(\sigma;\beta_1,...,\beta_\sigma)\cup\left\{0\right\},$
by the BCH bound, we can conclude that the minimum distance $d\ge \frac{2^{2p}-4}{2p}+\frac{2^p-2p-2}{4p}+1.$
\end{IEEEproof}

Several lower bounds on the minimum distance of binary cyclic codes are given in Table \ref{tab3}.
Compared with the square-root bound, the lower bound on the minimum distance of the cyclic codes $C$ in Theorem \ref{thm1} is much larger.
\begin{table}[ht]
            \renewcommand{\arraystretch}{1.5}
            \centering
            \caption{Comparison between the lower bounds on minimum distance in Theorem \ref{thm1} and the square-root bound \centering}
           \label{tab3}
            \begin{tabular}{|c|c|c|c|}
                \hline
                $p$&$n=2^{2p}-1$&$d(C)$&Square-root bound\\
                \hline
                 3&63&$\ge10$&8\\
                \hline
                 5&1023&$\ge103$&32\\
                \hline
                7&16383&$\ge1174$&128\\
                \hline
                11&4194303&$\ge190696$&2048\\
              \hline

            \end{tabular}
        \end{table}
        
In the proof of Theorem \ref{thm1}, for the case 
$\sigma_0<\frac{2^{2p}-2^{p+1}}{4p}$, we can choose suitable $\beta_{\sigma_0+1},...,\beta_{\frac{2^{2p}-2^{p+1}}{4p}}$ such that the dual code of the code constructed in Theorem \ref{thm1} has good dual distance. Specifically, we have following Theorem.
\begin{theorem}\label{thm2}
    Let $p$ be an odd prime and $n=2^{2p}-1$. Then, there exist integers $\sigma$ and $\beta_1, \beta_2, \cdots,\beta_\sigma$ satisfying $\frac{2^{2p}-2^{p+1}}{4p}\le\sigma\le\frac{2^{2p}-4}{4p}$ and
$1\le\beta_1<\beta_2<\cdots<\beta_\sigma\le\frac{2^{2p}-2^p-2}{2p}$ such that the dual code $C^\perp$ has parameters $[2^{2p}-1,2^{2p-1}-1,d^\perp\ge2^{\lfloor\log_22p\rfloor+1}-1]$, where the code $C$ is the binary cyclic code with defining set $Z_1(\sigma;\beta_1,...,\beta_\sigma)\cup\left\{0\right\}$, as constructed in Theorem \ref{thm1}. 
\end{theorem}
\begin{IEEEproof}
Let $D_0=\left\{1,2,...,\frac{2^{2p}-4}{2p},\frac{2^{2p}-4}{2p}+1,...,\frac{2^{2p}-4}{2p}+\frac{2^p-2p-2}{4p}-1\right\}$ given in the proof of Theorem \ref{thm1}. Note that 
\begin{align*}
       \frac{2^{2p+1}}{2^{\lfloor\log_24p\rfloor}}-\frac{2^{2p}-4}{2p}-\frac{2^p-2p-2}{4p}\ge&\frac{2^{2p+1}}{4p-1}-\frac{2^{2p}-4}{2p}-\frac{2^p-2p-2}{4p}\\
      =&\frac{2^{2p}+4(4p-1)}{2p(4p-1)}-\frac{2^p-2p-2}{4p}>0,
   \end{align*}
   then for any $1\le i\le \frac{2^{2p}-4}{2p}+\frac{2^p-2p-2}{4p}-1$, we have $i\le2^{2p+1-\lfloor\log_24p\rfloor}-2$, which implies that $wt(\pi(i))\le 2p-\lfloor\log_24p\rfloor.$ By Corollary \ref{cor1}, it holds that 
   \[wt(\pi(x))\le2p-\lfloor\log_24p\rfloor \textnormal{ for any }x\in\bigcup_{x\in D_0} C_{x}^{(2,n)}.\]
Note that for any $1< j\le 2^{\lfloor\log_22p\rfloor+1}-1$, we have $wt(\pi(2^{2p}-j))\geq 2p-\lfloor\log_22p\rfloor>2p-\lfloor\log_24p\rfloor$. By the above, we obtain that
\[2^{2p}-j\not\in \bigcup_{x\in D_0} C_{x}^{(2,n)}.\]  Furthermore, since $\pi(2^{2p}-j)=(\bm{1}_{2p-1-\lfloor\log_22p\rfloor},\bm{a})$ with some ${\bm{a}}\in\mathbb{F}_2^{\lfloor\log_22p\rfloor+1},$ by Corollary \ref{cor3}, we can conclude that $\left|C_{2^{2p}-j}^{(2,n)}\right|=2p$ since $2p-1-\lfloor\log_22p\rfloor\ge p$. 

Let $\sigma_0$ be the integer given in the proof of Theorem \ref{thm1}, i.e., the smallest integer such that $\left\{x:x\in D_0, |C_x^{(2,n)}|=2p\right\}\subseteq\bigcup_{\beta=1}^{\sigma_0}C_{j_\beta}^{(2,n)}$. 

When $\frac{2^{2p}-2^{p+1}}{4p}\le\sigma_0\le\frac{2^{2p}-4}{4p}$, from Eq. \eqref{6}, we have $\left|C_x^{(2,n)}\right|=p$ for any $x\in Z_1(\sigma_0; 1,2,...,\sigma_0)\backslash(\bigcup_{x\in D_0}C_x^{(2,n)})$.
Thus $2^{2p}-j\notin Z_1(\sigma_0;1,2,...,\sigma_0)\backslash(\bigcup_{x\in D_0}C_x^{(2,n)})$, hence $2^{2p}-j\notin Z_1(\sigma_0;1,2,...,\sigma_0)$.

When $\sigma_0<\frac{2^{2p}-2^{p+1}}{4p}$, note that  
\begin{align*}
    &\left|\left\{2^{2p}-j: j=2,...,2^{\lfloor\log_22p\rfloor+1}-1\right\}_{odd}\right|=2^{\lfloor\log_22p\rfloor}-1<\frac{2^{2p}-2^p-2}{2p}-\frac{2^{2p}-2^{p+1}}{4p},
\end{align*}
then we can always choose suitable $\sigma_0 <\beta_{\sigma_0+1}<\beta_{\sigma_0+2}<\cdots<\beta_{\frac{2^{2p}-2^{p+1}}{4p}}\leq \frac{2^{2p}-2^p-2}{2p}$ such that $2^{2p}-j\notin\bigcup_{l=\sigma_0+1}^{\frac{2^{2p}-2^{p+1}}{4p}}C_{\beta_l}^{(2,n)}$ for any $1< j\le 2^{\lfloor\log_22p\rfloor+1}-1$. Thus, for any $1< j\le 2^{\lfloor\log_22p\rfloor+1}-1$, 
\begin{align*}
2^{2p}j\notin&\left(\bigcup_{\beta=1}^{\sigma_0}C_{j_\beta}^{(2,n)}\right)\bigcup\left(\bigcup_{l=\sigma_0+1}^{\frac{2^{2p}-2^{p+1}}{4p}}C_{j_{\beta_l}}^{(2,n)}\right)\bigcup\left(\bigcup_{\alpha=1}^{\frac{2^p-2}{p}}C_{i_\alpha}^{(2,n)}\right)\\=&Z_1\left(\frac{2^{2p}-2^{p+1}}{4p};1,2,...,\sigma_0,\beta_{\sigma_0+1},...,\beta_{\frac{2^{2p}-2^{p+1}}{4p}}\right).
\end{align*}

In summary,  for any $2^{2p}-2^{\lfloor\log_22p\rfloor+1}+1\le x< 2^{2p}-1$, we always have $x\not\in Z_1(\sigma;\beta_1,...,\beta_\sigma)\cup \{0\}$, where $Z_1(\sigma;\beta_1,...,\beta_\sigma)\cup \{0\}$ is the defining set of the binary cyclic codes constructed in Theorem \ref{thm1}.  Hence $x \in \mathbb{Z}_n\backslash(Z_1(\sigma;\beta_1,...,\beta_\sigma)\cup\left\{0\right\})$, which implies that
$\left\{1,...,2^{\lfloor\log_22p\rfloor+1}-2\right\}\subseteq -(\mathbb{Z}_n\backslash(Z_1(\sigma;\beta_1,...,\beta_\sigma)\cup\left\{0\right\}))$. By the BCH bound, the minimum distance $d^\perp$ of the dual code $C^{\perp}$ satisfies $d^\perp\ge2^{\lfloor\log_22p\rfloor+1}-1.$
\end{IEEEproof}

\subsection{\texorpdfstring{$m=p_1p_2$}{m=p1p2}}
In this subsection, we suppose $m=p_1p_2$, where $p_2>p_1$ are two odd primes. Similarly, by Corollary \ref{cor3}, we can obtain that 
$$\left|\left\{C_x^{(2,n)}:\left|C_x^{(2,n)}\right|=p_1\right\}\right|=\frac{2^{p_1}-2}{p_1},$$
$$\left|\left\{C_x^{(2,n)}:\left|C_x^{(2,n)}\right|=p_2\right\}\right|=\frac{2^{p_2}-2}{p_2}$$
and
$$\left|\left\{C_x^{(2,n)}:\left|C_x^{(2,n)}\right|=p_1p_2\right\}\right|=\frac{2^{p_1p_2}-2^{p_1}-2^{p_2}+2}{p_1p_2}.$$
Let $1\le i_1<i_2<\cdots<i_{\frac{2^{p_1}-2}{p_1}}\le n-1$ be the coset leaders of the 2-cyclotomic cosets $C_{i_\alpha}^{(2,n)}$ with $|C_{i_\alpha}^{(2,n)}|=p_1(1\le\alpha\le\frac{2^{p_1}-2}{p_1})$, $1\le j_1<j_2<\cdots<j_{\frac{2^{p_2}-2}{p_2}}\le n-1$ be the coset leaders of the 2-cyclotomic cosets $C_{j_\beta}^{(2,n)}$ with $|C_{j_\beta}^{(2,n)}|=p_2$ $(1\le \beta\le\frac{2^{p_2}-2}{p_2})$ and $1\le k_1<k_2<\cdots<k_{\frac{2^{p_1p_2}-2^{p_1}-2^{p_2}+2}{p_1p_2}}$ be the coset leaders of the 2-cyclotomic cosets $C_{i_\gamma}^{(2,n)}$ with $|C_{i_\alpha}^{(2,n)}|=p_1p_2$ $(1\le\gamma\le\frac{2^{p_1p_2}-2^{p_1}-2^{p_2}+2}{p_1p_2})$.

Let $\sigma$ be an integer with $\frac{2^{p_1p_2}-2^{p_1}-2^{p_2+1}+4}{2p_1p_2}\le\sigma\le\frac{2^{p_1p_2}-2^{p_1}}{2p_1p_2}<\frac{2^{p_1p_2}-2^{p_1}-2^{p_2}+2}{p_1p_2}$, then $0\le\frac{2^{p_1p_2}-2^{p_1}}{2p_2}-p_1\sigma\le\frac{2^{p_2}-2}{p_2}.$ For any $1\le\gamma_1<\gamma_2<\cdots<\gamma_\sigma\le\frac{2^{p_1p_2}-2^{p_1}-2^{p_2}+2}{p_1p_2}$, we define

\begin{align*}
    Z_2(\sigma;\gamma_1,\gamma_2,...,\gamma_\sigma)=&\left(\bigcup_{\alpha=1}^{\frac{2^{p_1}-2}{2p_1}}C_{i_\alpha}^{(2,n)}\right)\bigcup\left(\bigcup_{l=1}^{\sigma}C_{k_{\gamma_l}}^{(2,n)}\right)\bigcup\left(\bigcup_{\beta=1}^{\frac{2^{p_1p_2}-2^{p_1}}{2p_2}-p_1\sigma}C_{j_\beta}^{(2,n)}\right).
\end{align*}
  Then 
  \begin{align*}
      \left|Z_2(\sigma;\gamma_1,\gamma_2,...,\gamma_\sigma)\right|=&p_1\cdot\frac{2^{p_1}-2}{2p_1}+p_2\cdot\left(\frac{2^{p_1p_2}-2^{p_1}}{2p_2}-p_1\sigma\right)+p_1p_2\cdot\sigma=2^{p_1p_2-1}-1.
  \end{align*}

Thus, we can get a binary $[2^m-1,2^{m-1}]$ cyclic code in the following Theorem.

\begin{theorem}\label{thm5}
Suppose $p_1<p_2$ are two odd primes. Let $m=p_1p_2$, then there exists a binary $[2^m-1,2^{m-1},d]$-cyclic code $C$ with $d\ge\left\lfloor\frac{2^{p_1p_2}-1}{p_1p_2}\right\rfloor+\left\lfloor\frac{2^{p_1}-1}{2p_1p_2}\right\rfloor+\left\lfloor\frac{2^{p_2}-1}{2p_1p_2}\right\rfloor+\left\lfloor\frac{2^{p_1}-1}{4p_1p_2}\right\rfloor+\left\lfloor\frac{2^{p_2}-1}{4p_1p_2}\right\rfloor.$

\end{theorem}

\begin{IEEEproof}Define the set
  \begin{align*}
      D_0=&\left\{1,2,...,\left\lfloor\frac{2^{p_1p_2}-1}{p_1p_2}\right\rfloor,...,\left\lfloor\frac{2^{p_1p_2}-1}{p_1p_2}\right\rfloor+\left\lfloor\frac{2^{p_1}-1}{2p_1p_2}\right\rfloor+\left\lfloor\frac{2^{p_2}-1}{2p_1p_2}\right\rfloor+\left\lfloor\frac{2^{p_1}-1}{4p_1p_2}\right\rfloor+\left\lfloor\frac{2^{p_2}-1}{4p_1p_2}\right\rfloor-1\right\}.
  \end{align*}
    Since
\begin{align*}
    &\left\lfloor\frac{2^{p_1}-1}{p_1p_2}\right\rfloor\sum_{i=0}^{p_2-1}2^{ip_1}\le\left\lfloor\frac{2^{p_1p_2}-1}{p_1p_2}\right\rfloor
  \le\left\lfloor\frac{2^{p_1p_2}-1}{p_1p_2}\right\rfloor+\left\lfloor\frac{2^{p_1}-1}{2p_1p_2}\right\rfloor+\left\lfloor\frac{2^{p_2}-1}{2p_1p_2}\right\rfloor+\left\lfloor\frac{2^{p_1}-1}{4p_1p_2}\right\rfloor+\left\lfloor\frac{2^{p_2}-1}{4p_1p_2}\right\rfloor-1
\end{align*}
    and
\begin{align*}
    &\left\lfloor\frac{2^{p_2}-1}{p_1p_2}\right\rfloor\sum_{i=0}^{p_1-1}2^{ip_2}\le\left\lfloor\frac{2^{p_1p_2}-1}{p_1p_2}\right\rfloor
  \le\left\lfloor\frac{2^{p_1p_2}-1}{p_1p_2}\right\rfloor+\left\lfloor\frac{2^{p_1}-1}{2p_1p_2}\right\rfloor+\left\lfloor\frac{2^{p_2}-1}{2p_1p_2}\right\rfloor+\left\lfloor\frac{2^{p_1}-1}{4p_1p_2}\right\rfloor+\left\lfloor\frac{2^{p_2}-1}{4p_1p_2}\right\rfloor-1,
\end{align*}
we can obtain that 
\begin{align*}
D_1&\triangleq\left\{a\cdot\sum_{i=0}^{p_1-1}2^{ip_2}:a=1,2,...,\left\lfloor\frac{2^{p_2}-1}{p_1p_2}\right\rfloor\right\}\subseteq \left\{x:x\in D_0,\left|C_x^{(2,n)}\right|=p_2\right\}    
\end{align*}
and
\begin{align*}
    D_2&\triangleq\left\{b\cdot\sum_{i=0}^{p_2-1}2^{ip_1}:b=1,2,...,\left\lfloor\frac{2^{p_1}-1}{p_1p_2}\right\rfloor\right\}\subseteq \left\{x:x\in D_0,\left|C_x^{(2,n)}\right|=p_1\right\}.
\end{align*}
 Moreover, since
 \begin{align*}
     \left(\frac{2^{p_1}-1}{p_1p_2}+1\right)\sum_{i=0}^{p_2-1}2^{ip_1}&=\frac{2^{p_1p_2}-1}{p_1p_2}+\sum_{i=0}^{p_2-1}2^{ip_1} > \frac{2^{p_1p_2}-1}{p_1p_2}+2^{(p_2-1)p_1}+2^{p_1}\\
     &\geq\left\lfloor\frac{2^{p_1p_2}-1}{p_1p_2}\right\rfloor+\left\lfloor\frac{2^{p_1}-1}{2p_1p_2}\right\rfloor+\left\lfloor\frac{2^{p_2}-1}{2p_1p_2}\right\rfloor+\left\lfloor\frac{2^{p_1}-1}{4p_1p_2}\right\rfloor+\left\lfloor\frac{2^{p_2}-1}{4p_1p_2}\right\rfloor-1,
 \end{align*}
we have
 $$\left|\left\{x:x\in D_0,\left|C_x^{(2,n)}\right|=p_1\right\}\right|\leq\frac{2^{p_1}-1}{p_1p_2}+1\le\frac{2^{p_1}-2}{2p_1}.$$
Therefore,
$$\left\{x:x\in\bigcup_{x\in D_0}C_x^{(2,n)},\left|C_x^{(2,n)}\right|=p_1\right\}\subseteq \bigcup_{\alpha=1}^{\frac{2^{p_1}-2}{2p_1}} C_{i_\alpha}^{(2,n)}.$$
Note that 
$$\left|(D_1)_{odd}\right|=\frac{1}{2}\left\lfloor\left(\left\lfloor\frac{2^{p_2}-1}{p_1p_2}\right\rfloor+1\right)\right\rfloor,\left|(D_2)_{odd}\right|=\frac{1}{2}\left\lfloor\left(\left\lfloor\frac{2^{p_1}-1}{p_1p_2}\right\rfloor+1\right)\right\rfloor,$$
and 
\begin{align*}
    \left|(D_0\backslash(D_1\cup D_2))_{odd}\right|=&\frac{1}{2}\left\lfloor\left(\left\lfloor\frac{2^{p_1p_2}-1}{p_1p_2}\right\rfloor+\left\lfloor\frac{2^{p_1}-1}{2p_1p_2}\right\rfloor+\left\lfloor\frac{2^{p_2}-1}{2p_1p_2}\right\rfloor\right.\right.+\left.\left.\left\lfloor\frac{2^{p_1}-1}{4p_1p_2}\right\rfloor+\left\lfloor\frac{2^{p_2}-1}{4p_1p_2}\right\rfloor\right)\right\rfloor\\
    &-\frac{1}{2}\left\lfloor\left(\left\lfloor\frac{2^{p_1}-1}{p_1p_2}\right\rfloor+1\right)\right\rfloor
    -\frac{1}{2}\left\lfloor\left(\left\lfloor\frac{2^{p_2}-1}{p_1p_2}\right\rfloor+1\right)\right\rfloor.
\end{align*}
Thus, we can conclude that 
\begin{align*}
          \left|\bigcup_{i\in D_0}C_i^{(2,n)}\right|
          \le& p_1p_2\left[\frac{1}{2}\left\lfloor\left(\left\lfloor\frac{2^{p_1p_2}-1}{p_1p_2}\right\rfloor+\left\lfloor\frac{2^{p_2}-1}{2p_1p_2}\right\rfloor\right.\right.\right.
          +\left.\left.\left.\left\lfloor\frac{2^{p_1}-1}{2p_1p_2}\right\rfloor+\left\lfloor\frac{2^{p_1}-1}{4p_1p_2}\right\rfloor+\left\lfloor\frac{2^{p_2}-1}{4p_1p_2}\right\rfloor\right)\right\rfloor\right]
          \\&-\frac{1}{2}p_1p_2\cdot\left\lfloor\left(\left\lfloor\frac{2^{p_1}-1}{p_1p_2}\right\rfloor+1\right)\right\rfloor
          -\frac{1}{2}p_1p_2\cdot\left\lfloor\left(\left\lfloor\frac{2^{p_2}-1}{p_1p_2}\right\rfloor+1\right)\right\rfloor\\
          &+p_2 \cdot\frac{1}{2}\left\lfloor\left(\left\lfloor\frac{2^{p_2}-1}{p_1p_2}\right\rfloor+1\right)\right\rfloor+p_1\cdot\frac{1}{2}\left\lfloor\left(\left\lfloor\frac{2^{p_1}-1}{p_1p_2}\right\rfloor+1\right)\right\rfloor\\
          \le &p_1p_2\cdot\frac{2^{p_1p_2}-1}{2p_1p_2}
          +\frac{p_1p_2}{2}\cdot\left\lfloor\frac{2^{p_2}-1}{2p_1p_2}\right\rfloor
          +\frac{p_1p_2}{2}\cdot\left\lfloor\frac{2^{p_1}-1}{2p_1p_2}\right\rfloor
          +\frac{p_1p_2}{2}\cdot\left\lfloor\frac{2^{p_2}-1}{4p_1p_2}\right\rfloor
          +\frac{p_1p_2}{2}\cdot\left\lfloor\frac{2^{p_1}-1}{4p_1p_2}\right\rfloor\\
          &-\frac{p_1p_2-p_1}{2}\cdot\left\lfloor\left(\left\lfloor\frac{2^{p_1}-1}{p_1p_2}\right\rfloor+1\right)\right\rfloor
          -\frac{p_1p_2-p_2}{2}\cdot\left\lfloor\left(\left\lfloor\frac{2^{p_2}-1}{p_1p_2}\right\rfloor+1\right)\right\rfloor\\
          \le&\frac{2^{p_1p_2}-1}{2}+p_1\left(\frac{p_2}{2}\cdot\left\lfloor\frac{2^{p_1}-1}{2p_1p_2}\right\rfloor\right.
          +\left.\frac{p_2}{2}\cdot\left\lfloor\frac{2^{p_1}-1}{4p_1p_2}\right\rfloor
          -\frac{p_2-1}{2}\cdot\left\lfloor\frac{2^{p_1}-1}{p_1p_2}\right\rfloor\right)\\
          &+p_2\left(\frac{p_1}{2}\cdot\left\lfloor\frac{2^{p_2}-1}{2p_1p_2}\right\rfloor+\frac{p_1}{2}\cdot\left\lfloor\frac{2^{p_2}-1}{4p_1p_2}\right\rfloor\right.
          -\left.\frac{p_1-1}{2}\cdot\left\lfloor\frac{2^{p_2}-1}{p_1p_2}\right\rfloor\right)\\
          \le&\frac{2^{p_1p_2}-1}{2}+p_1\cdot\left\lfloor\frac{2^{p_1}-1}{p_1p_2}\right\rfloor\left(\frac{p_2}{4}+\frac{p_2}{8}\right.
           -\left.\frac{p_2-1}{2}\right)
          +p_2\cdot\left\lfloor\frac{2^{p_2}-1}{p_1p_2}\right\rfloor\left(\frac{p_1}{4}+\frac{p_1}{8}\right.-\left.\frac{p_1-1}{2}\right)\\
          <&\frac{2^{p_1p_2}-1}{2}.
    \end{align*}
Then $\left|\bigcup_{i\in D_0}C_i^{(2,n)}\right|\le\frac{2^{p_1p_2}-1}{2}$,
Let $\sigma_0$ be the smallest integer such that $$\left\{x:x\in D_0, \left|C_x^{(2,n)}\right|=p_1p_2\right\}\subseteq\bigcup_{\gamma=1}^{\sigma_0}C_{k_\gamma}^{(2,n)},$$
then $\left\{x:x\in \bigcup_{x\in D_0}C_{x}^{(2,n)},  \left|C_x^{(2,n)}\right|=p_1p_2\right\}=\bigcup_{\gamma=1}^{\sigma_0}C_{k_\gamma}^{(2,n)}$. By the above, $\sigma_0<\frac{2^{p_1p_2}-2^{p_1}}{2p_1p_2}$.

If $\frac{2^{p_1p_2}-2^{p_1}-2^{p_2+1}+4}{2p_1p_2}\le\sigma_0<\frac{2^{p_1p_2}-2^{p_1}}{2p_1p_2}$, then
\begin{equation}\label{9}
    \begin{aligned}
    \bigcup_{x\in D_0} C_x^{(2,n)}=&\left\{x:x\in \bigcup_{x\in D_0}C_{x}^{(2,n)},  \left|C_x^{(2,n)}\right|=p_1\right\}
\bigcup\left\{x:x\in \bigcup_{x\in D_0}C_{x}^{(2,n)}, \left|C_x^{(2,n)}\right|=p_2\right\}\\
&\bigcup\left\{x:x\in \bigcup_{x\in D_0}C_{x}^{(2,n)}, \left|C_x^{(2,n)}\right|=p_1p_2\right\}\\
    \subseteq&\left(\bigcup_{\alpha=1}^{\frac{2^{p_1}-2}{2p_1}}C_{i_\alpha}^{(2,n)}\right)\bigcup\left(\bigcup_{l=1}^{\sigma_0}C_{k_{l}}^{(2,n)}\right)
    \bigcup\left\{x:x\in \bigcup_{x\in D_0}C_{x}^{(2,n)},  \left|C_x^{(2,n)}\right|=p_2\right\}\\
\stackrel{\rm (b)}{\subseteq}&\left(\bigcup_{\alpha=1}^{\frac{2^{p_1}-2}{2p_1}}C_{i_\alpha}^{(2,n)}\right)\bigcup\left(\bigcup_{\beta=1}^{\frac{2^{p_1p_2}-2^{p_1}}{2p_2}-p_1\sigma_0}C_{j_\beta}^{(2,n)}\right)
\bigcup\left(\bigcup_{l=1}^{\sigma_0}C_{k_{l}}^{(2,n)}\right)\\
=&Z_2(\sigma_0;1,2,...,\sigma_0),\\
\end{aligned}
\end{equation}
where $(b)$ follows from $\left|\left(\bigcup_{\alpha=1}^{\frac{2^{p_1}-2}{2p_1}}C_{i_\alpha}^{(2,n)}\right)\bigcup\left(\bigcup_{x\in D_0}C_x^{(2,n)}\right)\right|<\frac{2^{p_1p_2}-2^{p_1}}{2}+p_1\cdot\frac{2^{p_1}-2}{2p_1}=\frac{2^{p_1p_2}-2}{2}.$

If $\sigma_0<\frac{2^{p_1p_2}-2^{p_1}-2^{p_2+1}+4}{2p_1p_2}$, for any $\sigma_0<\gamma_{\sigma_0+1}<\gamma_{\sigma_0+2}<\cdots<\gamma_{\frac{2^{p_1p_2}-2^{p_1}-2^{p_2+1}+4}{2p_1p_2}}\le\frac{2^{p_1p_2}-2^{p_1}-2^{p_2}+2}{p_1p_2}$, 

    \begin{align*}
       \bigcup_{x\in D_0} C_x^{(2,n)}
       =&\left\{x:x\in\bigcup_{x\in D_0}C_{x}^{(2,n)}, \left|C_x^{(2,n)}\right|=p_1\right\}
       \bigcup\left\{x:x\in \bigcup_{x\in D_0}C_{x}^{(2,n)}, \left|C_x^{(2,n)}\right|=p_2\right\}
       \\
       &\bigcup\left\{x:x\in \bigcup_{x\in D_0}C_{x}^{(2,n)}, \left|C_x^{(2,n)}\right|=p_1p_2\right\}\\
        \subseteq&\left(\bigcup_{\alpha=1}^{\frac{2^{p_1}-2}{2p_1}}C_{i_\alpha}^{(2,n)}\right)\bigcup\left(\bigcup_{l=1}^{\sigma_0}C_{k_{l}}^{(2,n)}\right)
        \bigcup\left\{x:x\in \bigcup_{x\in D_0}C_{x}^{(2,n)}, \left|C_x^{(2,n)}\right|=p_2\right\}\\
        \subseteq&\left(\bigcup_{\alpha=1}^{\frac{2^{p_1}-2}{2p_1}}C_{i_\alpha}^{(2,n)}\right)\bigcup\left(\bigcup_{\beta=1}^{\frac{2^{p_2}-2}{p_2}}C_{j_\beta}^{(2,n)}\right)
        \bigcup\left(\bigcup_{l=\sigma_0+1}^{\frac{2^{p_1p_2}-2^{p_1}-2^{p_2+1}+4}{2p_1p_2}}C_{k_{\gamma_l}}^{(2,n)}\right)
        \bigcup\left(\bigcup_{l=1}^{\sigma_0}C_{k_{l}}^{(2,n)}\right)\\
        =&Z_2\left(\frac{2^{p_1p_2}-2^{p_1}-2^{p_2+1}+4}{2p_1p_2};1,2,...,\sigma_0,\gamma_{\sigma_0+1},...,\gamma_{\frac{2^{p_1p_2}-2^{p_1}-2^{p_2+1}+4}{2p_1p_2}}\right).
    \end{align*}

Thus, we can deduce that there always exits an integer $\sigma$ with $\frac{2^{p_1p_2}-2^{p_1}-2^{p_2+1}+4}{2p_1p_2}\le\sigma<\frac{2^{p_1p_2}-2^{p_1}}{2p_1p_2}$ and $1 \leq \gamma_1<\gamma_2<...<\gamma_\sigma \leq \frac{2^{p_1p_2}-2^{p_1}-2^{p_2}+2}{p_1p_2}$ such that for any $1\le x\le\left\lfloor\frac{2^{p_1p_2}-1}{p_1p_2}\right\rfloor+\left\lfloor\frac{2^{p_1}-1}{2p_1p_2}\right\rfloor+\left\lfloor\frac{2^{p_2}-1}{2p_1p_2}\right\rfloor+\left\lfloor\frac{2^{p_1}-1}{4p_1p_2}\right\rfloor+\left\lfloor\frac{2^{p_2}-1}{4p_1p_2}\right\rfloor-1$, $x\in Z_2\left(\sigma;\gamma_1,\gamma_2,...,\gamma_\sigma\right).$ 
Let $C$ be the binary cyclic code with defining set $Z_2(\sigma;\gamma_1,\gamma_2,...,\gamma_\sigma)$, then $\dim(C)=n-|Z_2(\sigma;\gamma_1,\gamma_2,...,\gamma_\sigma)|=2^{m-1}$ and the minimum distance $d\ge\left\lfloor\frac{2^{p_1p_2}-1}{p_1p_2}\right\rfloor+\left\lfloor\frac{2^{p_1}-1}{2p_1p_2}\right\rfloor+\left\lfloor\frac{2^{p_2}-1}{2p_1p_2}\right\rfloor+\left\lfloor\frac{2^{p_1}-1}{4p_1p_2}\right\rfloor+\left\lfloor\frac{2^{p_2}-1}{4p_1p_2}\right\rfloor$ by the BCH bound.
\end{IEEEproof}
Several suitable pairs of $p_1$ and $p_2$, together with the corresponding lower bounds on the minimum distance of the binary cyclic codes, are listed in Table \ref{tab4}. Compared to the punctured binary Reed-Muller codes of order $(m-1)/2$, the lower bound on the minimum distance of the binary cyclic codes constructed in Theorem \ref{thm5} is significantly larger.
\begin{table*}[ht]
        \renewcommand{\arraystretch}{1.5}
        \centering
        \caption{Comparison between the lower bounds on minimum distance in Theorem \ref{thm5} and  punctured binary Reed-Muller codes\centering}
           \label{tab4}
            \begin{tabular}{|c|c|c|c|c|}
                \hline
                $p_1$&$p_2$&$n=2^{p_1p_2}-1$&$d(C)$&$d(PRM)=2^{(p_1p_2+1)/2}-1$\\
                \hline
                 3&5&32767&$\ge2185$&255\\
                \hline
                 3&7&2097151&$\ge99868$&2047\\
                \hline
                3&13&549755813887&$\ge14096303077$&1048575\\
                \hline
                5&13&36893488147419103231&$\ge 567592125344909374$&8589934591\\
                \hline 
        \end{tabular}
\end{table*}

\begin{remark}
    In \cite{ref7}, Sun et al. presented a family of binary cyclic codes with parameters $[n,(n+1)/2,d\ge\frac{n}{\log_2n}]$, whose lower bound is significantly larger than the square-root bound. However, the lower bound on the minimum distance of the two families of cyclic codes constructed in our Theorems \ref{thm1} and \ref{thm5} is even larger than $\frac{n}{\log_2n}.$
\end{remark}

Similar to Theorem \ref{thm2}, we present the parameters of the dual code constructed in Theorem \ref{thm5} and refer to Appendix \ref{B} for its proof.

\begin{theorem}\label{thm6}
   The dual code $C^\perp$ of the binary cyclic code constructed in Theorem \ref{thm5} has parameters $[2^{p_1p_2}-1,2^{p_1p_2-1}-1,d^\perp\ge 2^{\lfloor\log_2p_1p_2\rfloor+1}]$. 
\end{theorem}

\section{Third Construction: \texorpdfstring{$m$}{m} is odd}\label{set3}
Let $m$ be an odd integer and set $n=2^m-1$. Recently, considerable research has focused on the construction of binary $[2^m-1, 2^{m-1}, d]$ cyclic codes in this setting, as discussed in \cite{ref4}, \cite{ref6}, and \cite{ref5}. It is not difficult to observe that as the minimum distance $d$ increases, the dual distance $d^\perp$ tends to decrease correspondingly. In this section, we present two new families of binary $[2^m-1, 2^{m-1}]$ cyclic codes with good minimum distance $d$ and dual distance $d^\perp$. Notably, the lower bound of $d \cdot d^\perp$ can asymptotically approach $2n$.

\subsection{\texorpdfstring{$d\ge2^{(m+1)/2}-1$}{} and \texorpdfstring{$d^\perp\ge2^{(m+1)/2}$}{}}\label{V.A}
Let $n=2^m-1$, where $m$ is odd. For any $0\le x\le n-1$, let $\pi(x)=(x_0,x_1,...,x_{m-1})$, where $\pi$ is defined in Section \ref{set2}. Note that $wt(\pi(x))=m-wt(\pi(-x\mod n)),$ thus, $wt(\pi(x))\not=wt(\pi(-x\mod n))$ since $m$ is odd. So, we have
$$C_x^{(2,n)}\cap C_{-x}^{(2,n)}=\emptyset.$$
Let $1<b\le n-1,$ denote
$$T_{b}=\bigcup_{1\le i\le b}C_i^{(2,n)}.$$

For any subset $S\subseteq \mathbb{Z}_n$, define $$-S=\left\{-i\mod n\right\}\subseteq\mathbb{Z}_n,$$
then we have the following lemma.

\begin{lemma}\label{lem7}
$T_{2^{(m+1)/2}-2}\cap(-T_{2^{(m+1)/2}-2})=\emptyset.$
\end{lemma}

\begin{IEEEproof}
    Clearly, ${\rm{ord}}_n(2)=m$. If $T_{2^{(m+1)/2}-2}\cap
    (-T_{2^{(m+1)/2}-2})\not=\emptyset$, then
there exist integers $x, y$ and $t$ with $1\le x,y\le2^{(m+1)/2}-2$ and $0\le t\le m-1$, such that 
$$x\equiv-y\cdot2^t\pmod{2^m-1}.$$
When $(m+1)/2\le t <m$, we can derive that
$$y\equiv-x2^{m-t}\pmod{2^m-1}.$$
Thus, we may assume that $0\le t\le(m-1)/2.$
Then
$$1 \leq x+y2^{t}\le(2^{(m-1)/2}+1)\cdot(2^{(m+1)/2}-2)<2^m-1,$$ 
which contradicts to $x\equiv-y\cdot2^t\pmod{2^m-1}.$ We then conclude that $T_{2^{(m+1)/2}-2}\cap(-T_{2^{(m+1)/2}-2})=\emptyset$.
\end{IEEEproof}
Combining Lemma \ref{lem7} with the fact that $C_x^{(2,n)}\cap C_{-x}^{(2,n)}=\emptyset$, we derive that there exists a  subset $S_0\subseteq \Gamma_{(2,n)} \backslash\left\{0\right\}$ such that 
\begin{align*}
    &\left(\mathbb{Z}_n\backslash\left\{0\right\}\right)\backslash\left(T_{2^{(m+1)/2}-2}\bigcup \left(-T_{2^{(m+1)/2}-2}\right)\right)\\
    =&\bigcup_{i\in S_0}\left(C_{i}^{(2,n)}\bigcup C_{-i}^{(2,n)}\right).
\end{align*}
Define $$Z=T_{2^{(m+1)/2}-2}\bigcup\left(\bigcup_{i\in  S_0}C_i^{(2,n)}\right).$$

Clearly, $|Z|=2^{m-1}-1$. Then we present a construction of binary cyclic codes as follows.
\begin{theorem}\label{tm3}
Let $m$ be an odd integer and $n=2^m-1$. Then the binary cyclic code $C$ with defining set $Z$ has parameters $[2^m-1,2^{m-1},d\ge2^{(m+1)/2}-1]$ and the dual code $C^\perp$ has parameters $[2^m-1,2^{m-1}-1,d^\perp\ge2^{(m+1)/2}]$.

\end{theorem}

\begin{IEEEproof}
Clearly, $\dim(C)=n-(2^{m-1}-1)=2^{m-1}.$
    Note that $\left\{1,2,...,2^{(m+1)/2}-2\right\}\subseteq Z$, by the BCH bound, we obtain that $d(C)\ge2^{(m+1)/2}-1$. 
    
    The defining set of $C^{\perp}$ is $-(\mathbb{Z}_n\backslash Z)=Z\cup\left\{0\right\}$, we then deduce that  $\left\{0,1,2,...,2^{(m+1)/2}-2\right\}\subseteq Z\cup\left\{0\right\}$, which leads to $d(C^{\perp})\ge 2^{(m+1)/2}.$
\end{IEEEproof}

\begin{example}
    Let $m=5$ and $n=2^m-1=31$, then $(m+1)/2=3$. We can compute that 
    $$Z=\left\{1,2,3,4,5,6,8,9,10,12,16,17,18,20,24\right\}.$$

Then the binary cyclic code $C$ with defining set $Z$ has parameters $[31,16,7]$ and $C^{\perp}$ has parameters $[31,15,8]$, which has the best known parameters.
\end{example}

\begin{example}\label{exam1}
    Let $m=7$ and $n=2^m-1=127$, then $(m+1)/2=4$. We can compute that
\begin{align*}
    Z=&\left\{1,2,3,4,5,6,7,8,9,10,11,12,13,14,16,17,18,20,
   21,22,24,26,27,28,32,33,34,\right. \\&35,36,37,40,41,42,44,48,49,51,52,54,56,64,65,66,67,68,69,70,72,74,77,80,\\
   &\left.81,82,84,88,89,96,97,98,102,104,108,112\right\}.\\
   \end{align*}
     Then the binary cyclic code $C$ with defining set $Z$ has parameters $[127,64,19]$ and $C^{\perp}$ has parameters $[127,63,20].$
\end{example}

\begin{remark}
    In \cite{ref8}, the punctured binary Reed-Muller codes of order $(m-1)/2$ have parameters $[2^m-1,2^{m-1},2^{(m+1)/2}-1]$ when $m$ is odd while the minimum distance on the binary constructed in this section with length $n=2^m-1$ could be larger than $2^{(m+1)/2}-1$. Specially, in Example \ref{exam1}, the minimum distance of $C$ is $19$, which is larger than $2^{(m+1)/2}-1=15$.
\end{remark}

\subsection{\texorpdfstring{$d\ge 2^{\frac{m+3}{2}}-15$}{} and \texorpdfstring{$d^{\perp}\ge 2^{\frac{m-1}{2}}$}{} }\label{V.B}
    Sun et al. \cite{ref5} constructed several infinite families of binary cyclic codes with good minimum distance. In this section, we present a construction of binary cyclic codes that achieve a better lower bound on the minimum distance while maintaining the same lower bound on the dual minimum distance.

    Let $m > 1$ be a positive integer and let $n=2^m-1$. For each $i\in \{0,1\}$, denote 
    $$S_{(i,m)} = \{j:1\leq j \leq n-1, wt(j) \equiv i \pmod 2\}.$$
   Since $wt(x)$ is constant on each 2-cyclotomic coset $C^{(2,n)}_b$, the set $S_{(i,m)}$ must be a union of certain 2-cyclotomic cosets modulo $n$. To proceed, we require some known results on the structure of $S_{(i,m)}$ and the coset leaders of 2-cyclotomic cosets.

    \begin{lemma}[\hspace{-0.01em}\cite{ref4}]
        Let $S_{(i,m)}$ be defined as above. Then, we have
        \begin{itemize}
            \item [(1)] If $m\geq 7$ is odd, then $\left|S_{(0,m)}\right| = \left|S_{(1,m)}\right|  = 2^{m-1}-1.$ 
            \item [(2)] If $m\geq 6$ is even, then $\left|S_{(0,m)}\right| =2^{m-1}-2$ and $\left|S_{(1,m)}\right|  = 2^{m-1}.$ 
        \end{itemize}
    \end{lemma}

    \begin{lemma}[\hspace{-0.01em}\cite{ref10}]
        \label{Lem1} Let $m \geq 7$ be odd and $i$ be an integer with $1\leq i \leq 2^{\frac{m+3}{2}}-1$ and $i \equiv 1 \pmod 2$. Then, $\left|C^{\,(2,n)}_i \right|= m $ and $i$ is not a coset leader if and only if $i\in\left\{2^{\frac{m+1}{2}}+1,2^{\frac{m+1}{2}}+2^{\frac{m-1}{2}}+1 \right\}$.
    \end{lemma}
  The following two lemmas will be used to construct the desired defining set of our binary cyclic codes.
    \begin{lemma}
        \label{Lem2}
        Let $m\geq 7$ be odd. Then,
        \begin{align*}
            &\big(\bigcup_{i=1}^{4\cdot(2^{\frac{m-1}{2}}-4)}C_i^{\,(2,n)}\big) \bigcap \big(\bigcup_{j=1}^{4\cdot(2^{\frac{m-1}{2}}-4)}C_{n-j}^{\,(2,n)}\big) = C_{2^{\frac{m-1}{2}}-1}^{\,(2,n)} \cup C_{2^{\frac{m+1}{2}}-3}^{\,(2,n)} \cup C_{2^{\frac{m+1}{2}}-1}^{\,(2,n)} \cup C_{3\cdot2^{\frac{m-1}{2}}-1}^{\,(2,n)}.
        \end{align*}
    \end{lemma}
    \begin{IEEEproof}
        For $1 \leq i \leq 4(2^{\frac{m-1}{2}} - 4)$, we have $C_{i}^{(2,n)} \subseteq \bigcup_{j=1}^{4(2^{\frac{m-1}{2}} - 4)} C_{n-j}^{(2,n)}$ if and only if $i \in C_{n-j}^{(2,n)}$ for some $1 \leq j \leq 4(2^{\frac{m-1}{2}} - 4)$. This is equivalent to the existence of an integer $l$ with $0 \leq l \leq m-1$ such that
\begin{equation} \label{Equation8} 
i + j \cdot 2^l \equiv 0 \pmod{n}.
\end{equation}
We discuss the following cases separately.

\emph{Case 1}: If $0 \leq l \leq \frac{m-3}{2}$, then for any $1 \leq i, j \leq 4 \cdot (2^{\frac{m-1}{2}} - 4)$, it holds that 
\begin{equation*}
                        \begin{aligned}
                            i+j\cdot 2^l &< 8\cdot(2^{\frac{m-3}{2}}-1)(2^\frac{m-3}{2}+1) \\
                            &= 8 \cdot (2^{m-3} -1)< n,
                        \end{aligned}
                    \end{equation*}
implying that Eq.~\eqref{Equation8} has no solution in this case.

\emph{Case 2}: If $l = \frac{m-1}{2}$, then for any $1 \leq i, j \leq 4 \cdot (2^{\frac{m-1}{2}} - 4)$, we have 
\begin{equation*}
                        \begin{aligned}
                            i+j\cdot 2^l &\leq 2^4\cdot (2^{\frac{m-5}{2}}-1)(2^\frac{m-1}{2}+1) \\
                            &= 2^4 \cdot (2^{m-3} -2^{\frac{m-3}{2}}-2^{\frac{m-5}{2}} -1) < 2n.
                        \end{aligned}
                    \end{equation*}
In this case, Eq.~\eqref{Equation8} is equivalent to $i + j \cdot 2^{\frac{m-1}{2}} = 2^m - 1$. Thus, the solutions are given by $(i, j) = (2^{\frac{m-1}{2}} - 1, 2^{\frac{m+1}{2}} - 1)$, $(2^{\frac{m+1}{2}} - 1, 2^{\frac{m+1}{2}} - 2)$, and $(3 \cdot 2^{\frac{m-1}{2}} - 1, 2^{\frac{m+1}{2}} - 3)$.

\emph{Case 3}: If $l = \frac{m+1}{2}$, then $m - l = \frac{m-1}{2}$. Similar to Case 2, Eq.~\eqref{Equation8} has the solutions: $(i, j) = (2^{\frac{m+1}{2}} - 1, 2^{\frac{m-1}{2}} - 1)$, $(2^{\frac{m+1}{2}} - 2, 2^{\frac{m+1}{2}} - 1)$, and $(2^{\frac{m+1}{2}} - 3, 3 \cdot 2^{\frac{m-1}{2}} - 1)$.

\emph{Case 4}: If $\frac{m+3}{2} \leq l \leq m-1$, then $1 \leq m - l \leq \frac{m-3}{2}$. Similar to Case 1, Eq.~\eqref{Equation8} has no solution in this case.

In summary, we conclude that
        $\big(\cup_{i=1}^{4\cdot(2^{\frac{m-1}{2}}-4)}C_i^{\,(2,n)}\big) \bigcap \big(\cup_{j=1}^{4\cdot(2^{\frac{m-1}{2}}-4)}C_{n-j}^{\,(2,n)}\big) = C_{2^{\frac{m-1}{2}}-1}^{\,(2,n)} \cup C_{2^{\frac{m+1}{2}}-3}^{\,(2,n)} \cup C_{2^{\frac{m+1}{2}}-1}^{\,(2,n)} \cup C_{3\cdot2^{\frac{m-1}{2}}-1}^{\,(2,n)}.$ This completes the proof.
    \end{IEEEproof}

    \begin{lemma}
        \label{Lem3}   
        Let $m\geq 9$ be odd. Then for $j \in\{9,11,13,15\}$, it holds that 
        $$C_{2^{\frac{m+3}{2}}-j}^{\,(2,n)} \not\subseteq \bigcup_{i=1}^{4\cdot(2^{\frac{m-1}{2}}-4)}C_{n-i}^{\,(2,n)}.$$ 
    \end{lemma}
    \begin{IEEEproof}
    It is equivalent to prove that  $2^{\frac{m+3}{2}}-j\notin \bigcup_{i=1}^{4\cdot(2^{\frac{m-1}{2}}-4)}C_{n-i}^{\,(2,n)}$ for any $j\in \{9,11,13,15\}$. Suppose $2^{\frac{m+3}{2}}-j\in C_{n-i}^{\,(2,n)}$ for some $1\leq i\leq 4\cdot(2^{\frac{m-1}{2}}-4)$, then there exists an integer $\ell$ with $0\leq \ell \leq m-1$ such that 
        \begin{equation}
            \label{Equation11}
            (2^{\frac{m+3}{2}}-j)2^{\ell}\equiv n-i \pmod n. 
        \end{equation}
        
\emph{Case 1}: If $0\leq \ell \leq \frac{m-3}{2}$, then 
            \begin{eqnarray*}
                \begin{aligned}
                    (2^{\frac{m+3}{2}}-j)2^{\ell} \text{ mod } n &\leq (2^{\frac{m+3}{2}}-j)2^{ \frac{m-3}{2}}\\
                    &< 2^{m}-8\cdot2^{ \frac{m-3}{2}} \\
                    &< 2^{m}-1 -4\cdot(2^{\frac{m-1}{2}}-4)
                    \leq n-i,
                \end{aligned}
            \end{eqnarray*}
            which contradicts to Eq. \eqref{Equation11}.

\emph{Case 2}: If $\frac{m-1}{2}\leq \ell \leq m-4$, then 
                        \begin{align*}
                            (2^{\frac{m+3}{2}}-j)2^{\ell} \text{ mod } n &=2^m-1+2^{\ell-\frac{m-3}{2}} - j\cdot2^{\ell}\\
                            &\leq 2^m-1+2 - j\cdot2^{\frac{m-1}{2}}\\
                            & < 2^{m}+1-8\cdot2^{ \frac{m-3}{2}}\\
                            & < n-i,
                        \end{align*}
                       which also contradicts to Eq. \eqref{Equation11}.

 \emph{Case 3}: If $\ell = m-3$, then $(2^{\frac{m+3}{2}}-j)2^{m-3}  \equiv 2^{\frac{m-3}{2}} - j\cdot 2^{m-3} \pmod n$.
                \begin{itemize}
                    \item If $j=9$, then $(2^{\frac{m+3}{2}}-9)2^{m-3} \text{ mod } n = 7\cdot 2^{m-3} + 2^{\frac{m-3}{2}}-2$.
                    \item If $j=11$, then $(2^{\frac{m+3}{2}}-11)2^{m-3} \text{ mod } n = 5\cdot 2^{m-3} + 2^{\frac{m-3}{2}}-2$.
                    \item If $j=13$, then $(2^{\frac{m+3}{2}}-13)2^{m-3} \text{ mod } n = 3\cdot 2^{m-3} + 2^{\frac{m-3}{2}}-2$.
                    \item If $j=15$, then $(2^{\frac{m+3}{2}}-15)2^{m-3} \text{ mod } n = 2^{m-3} + 2^{\frac{m-3}{2}}-2$.
                \end{itemize}
                Therefore, $$(2^{\frac{m+3}{2}}-j)2^{m-3} \text{ mod } n \leq  7\cdot 2^{m-3} + 2^{\frac{m-3}{2}}-2  <n-i,$$
which also contradicts to Eq. \eqref{Equation11}.

 \emph{Case 4}: If $\ell = m-2$, then $(2^{\frac{m+3}{2}}-j)2^{m-2}  \equiv 2^{\frac{m-1}{2}} - j\cdot 2^{m-2} \pmod n$.
                \begin{itemize}
                    \item If $j=9$, then $(2^{\frac{m+3}{2}}-9)2^{m-2} \text{ mod } n = 3\cdot 2^{m-2} + 2^{\frac{m-1}{2}}-3$.
                    \item If $j=11$, then $(2^{\frac{m+3}{2}}-11)2^{m-2} \text{ mod } n = 2^{m-2} + 2^{\frac{m-1}{2}}-3$.
                    \item If $j=13$, then $(2^{\frac{m+3}{2}}-13)2^{m-2} \text{ mod } n = 3\cdot 2^{m-2} + 2^{\frac{m-1}{2}}-4$.
                    \item If $j=15$, then $(2^{\frac{m+3}{2}}-15)2^{m-2} \text{ mod } n = 2^{m-2} + 2^{\frac{m-1}{2}}-4$.
                \end{itemize}
                Therefore,
                \begin{align}
                    (2^{\frac{m+3}{2}}-j)2^{m-2} \text{ mod } n 
                    \leq  3\cdot 2^{m-2} + 2^{\frac{m-1}{2}}-3  
                    <n-i, \nonumber
                \end{align}
 which also contradicts to Eq. \eqref{Equation11}.
 
 \emph{Case 5}:  If $\ell = m-1$, then $(2^{\frac{m+3}{2}}-j)2^{m-1} \equiv 2^{\frac{m+1}{2}} - j\cdot 2^{m-1} \pmod n$.
                \begin{itemize}
                    \item If $j=9$, then $(2^{\frac{m+3}{2}}-9)2^{m-1} \text{ mod } n = 2^{m-1} + 2^{\frac{m+1}{2}}-5$.
                    \item If $j=11$, then $(2^{\frac{m+3}{2}}-11)2^{m-1} \text{ mod } n = 2^{m-1} + 2^{\frac{m+1}{2}}-6$.
                    \item If $j=13$, then $(2^{\frac{m+3}{2}}-13)2^{m-1} \text{ mod } n = 2^{m-1} + 2^{\frac{m+1}{2}}-7$.
                    \item If $j=15$, then $(2^{\frac{m+3}{2}}-15)2^{m-1} \text{ mod } n = 2^{m-1} + 2^{\frac{m+1}{2}}-8$.
                \end{itemize}
                Therefore, 
                \begin{align*}
                    (2^{\frac{m+3}{2}}-j)2^{m-1} \text{ mod } n \leq  2^{m-1} + 2^{\frac{m+1}{2}}-5  
                    <n-i,
                \end{align*}
                 which contradicts Eq.  (\ref{Equation11}).

        In summary, we have $2^{\frac{m+3}{2}}-j \not \in \bigcup_{i=1}^{4\cdot(2^{\frac{m-1}{2}}-4)}C_{n-i}^{\,(2,n)}$ for $j \in\{9,11,13,15\}$. This completes the proof.
    \end{IEEEproof}
    
    Let $m\geq 9$ be odd and denote
    \[A = \bigcup_{j=1}^{4\cdot(2^{\frac{m-1}{2}}-4)} C_j^{\,(2,n)}, ~B = \bigcup_{j=1}^{4\cdot(2^{\frac{m-1}{2}}-4)} C_{n-j}^{\,(2,n)},\]
    and
    \[ B_{(0,i,m)}=\bigcup_{\substack{j=1 \\ wt(j)\equiv i \pmod 2}}^{4\cdot(2^{\frac{m-1}{2}}-4)} C_{n-j}^{\,(2,n)} \textnormal{ for } i=0,1.\]
    It is clear that $B = B_{(0,0,m)} \cup B_{(0,1,m)}$. Let
   
            $$B_{(1,i,m)}
            =S_{(1\oplus i ,m)}\backslash \left[\big(B_{(0,i,m)}\backslash (C^{\,(2,n)}_{t_{(i,m)}}\cup C^{\,(2,n)}_{u_{(i,m)}})\big)\cup C^{\,(2,n)}_{s_{(i,m)}} \cup C^{\,(2,n)}_{v_{(i,m)}}\right],$$
     
    where $i \in \{0,1\}$ and $\oplus$ denotes the binary addition and 
    \begin{equation*}
        \begin{aligned}
            t_{(i,m)} &=
            \begin{cases}
                2^{\frac{m-1}{2}}-1 &\substack{\text{ if } i=0 \text{ and } m\equiv3\pmod 4,\\ \text{ or } i=1 \text{ and } m\equiv1\pmod 4,} \\
                2^{\frac{m+1}{2}}-1 &\text{ otherwise,}
            \end{cases}\\
            u_{(i,m)} &=
            \begin{cases}
                2^{\frac{m+1}{2}}-3 &\substack{\text{ if } i=0 \text{ and } m\equiv 3 \pmod 4,\\ \text{ or } i=1 \text{ and } m\equiv 1\pmod 4,} \\
                3\cdot2^{\frac{m-1}{2}}-1  &\text{ otherwise,}
            \end{cases}\\
            s_{(i,m)} &=
            \begin{cases}
                2^{\frac{m+3}{2}}-9 &\substack{\text{ if } i=0 \text{ and } m\equiv 1 \pmod 4,\\ \text{ or } i=1 \text{ and } m\equiv 3\pmod 4,} \\
                2^{\frac{m+3}{2}}-11 &\text{ otherwise,}
            \end{cases}\\
            v_{(i,m)} &=
            \begin{cases}
                2^{\frac{m+3}{2}}-15 &\substack{\text{ if } i=0 \text{ and } m\equiv 1 \pmod 4,\\ \text{ or } i=1 \text{ and } m\equiv 3\pmod 4,} \\
                2^{\frac{m+3}{2}}-13 &\text{ otherwise.}
            \end{cases}
        \end{aligned}
    \end{equation*}
    Clearly, $B_{(0,i,m)} \subseteq S_{(i\oplus1,m)}$ and $B_{(1,i,m)} \subseteq S_{(i\oplus1,m)}$. By Lemma \ref{Lem3}, $2^{\frac{m+3}{2}} -j \notin \big(B_{(0,0,m)} \cup B_{(0,1,m)} \big)$ for $j\in\{9,11,13,15\}$. Thus, $(C^{\,(2,n)}_{s_{(i,m)}}\cup C^{\,(2,n)}_{v_{(i,m)}}) \cap B_{(0,i,m)} =  \emptyset$.

    Let 
    \[T_{(1,i,m)} = A\cup B_{(1,i,m)},\] 
    then $T_{(1,i,m)}$ is the union of some 2-cyclotomic cosets modulo $n$. Let $C_{(1,i,m)}$ be the binary cyclic code of length $2^m-1$ with defining set $T_{(1,i,m)}$. The aim of this subsection is to determine the parameters of $C_{(1, i, m)}$. First, we will introduce the following lemmas, which play an important role in studying the parameters
of the codes.

    \begin{lemma}
        \label{Lem4}
        For each $i \in\{0, 1\}$, we have
        \begin{itemize}
            \item [(1)] $B_{(0,i,m)} \cap A = C^{\,(2,n)}_{t_{(i,m)}}\cup C^{\,(2,n)}_{u_{(i,m)}}$.
            \item [(2)] $C^{\,(2,n)}_{s_{(i,m)}}\cup C^{\,(2,n)}_{v_{(i,m)}} \subseteq S_{(i\oplus1,m)}$.
            \item [(3)] $A\cap B_{(1,i,m)} = A\cap S_{(i\oplus1,m)}$.
        \end{itemize}
    \end{lemma}
    \begin{IEEEproof}
        (1) Note that $B_{(0,0,m)} \cap B_{(0,1,m)} = \emptyset$. By Lemma \ref{Lem2}, 
        \begin{equation*}
            \begin{aligned}
                &\big(B_{(0,0,m)} \cup B_{(0,1,m)}\big) \cap A \\
                =& C^{\,(2,n)}_{2^{\frac{m-1}{2}}-1} \cup C^{\,(2,n)}_{2^{\frac{m+1}{2}}-1}\cup C^{\,(2,n)}_{2^{\frac{m+1}{2}}-1} \cup C^{\,(2,n)}_{3\cdot2^{\frac{m+1}{2}}-1}.
            \end{aligned}
        \end{equation*}

        Note that $wt(2^{\frac{m-1}{2}}-1) =\frac{m-1}{2}$, $wt(2^{\frac{m+1}{2}}-3) =\frac{m-1}{2}$, $wt(2^{\frac{m+1}{2}}-1) =\frac{m+1}{2}$ and $wt(3\cdot2^{\frac{m+1}{2}}-1) =\frac{m+1}{2}$. Thus,
        \begin{itemize}
            \item If $m\equiv 1 \pmod 4$, then $B_{(0,0,m)} \cap A = C^{\,(2,n)}_{2^{\frac{m+1}{2}}-1} \cup C^{\,(2,n)}_{3\cdot2^{\frac{m+1}{2}}-1} $ and $B_{(0,1,m)} \cap A = C^{\,(2,n)}_{2^{\frac{m-1}{2}}-1} \cup C^{\,(2,n)}_{2^{\frac{m+1}{2}}-3}.$
            \item If $m\equiv 3 \pmod 4$, then $B_{(0,0,m)} \cap A = C^{\,(2,n)}_{2^{\frac{m-1}{2}}-1} \cup C^{\,(2,n)}_{2^{\frac{m+1}{2}}-3}$ and $B_{(0,1,m)} \cap A =C^{\,(2,n)}_{2^{\frac{m+1}{2}}-1} \cup C^{\,(2,n)}_{3\cdot2^{\frac{m+1}{2}}-1}.$
        \end{itemize}
        The desired result follows.

        (2) Note that $wt(2^{\frac{m+3}{2}}-9) =\frac{m+1}{2}$, $wt(2^{\frac{m+3}{2}}-11) =\frac{m-1}{2}$, $wt(2^{\frac{m+3}{2}}-13) =\frac{m-1}{2}$ and $wt(2^{\frac{m+3}{2}}-15) =\frac{m-3}{2}$. Thus, 
        \begin{itemize}
            \item If $m\equiv 3 \pmod 4$, then $C^{\,(2,n)}_{2^{\frac{m+3}{2}}-9} \cup C^{\,(2,n)}_{2^{\frac{m+3}{2}}-15} \in S_{(1\oplus1,m)} $ and $C^{\,(2,n)}_{2^{\frac{m+3}{2}}-11} \cup C^{\,(2,n)}_{2^{\frac{m+3}{2}}-13} \in S_{(1\oplus0,m)}$.
            \item If $m\equiv 1 \pmod 4$, then $C^{\,(2,n)}_{2^{\frac{m+3}{2}}-9} \cup C^{\,(2,n)}_{2^{\frac{m+3}{2}}-15} \in S_{(1\oplus0,m)} $ and $C^{\,(2,n)}_{2^{\frac{m+3}{2}}-11} \cup C^{\,(2,n)}_{2^{\frac{m+3}{2}}-13} \in S_{(1\oplus1,m)}$.
        \end{itemize}
        The desired result follows.

        (3) By Lemma \ref{Lem1}, $2^{\frac{m+3}{2}}-9$, $2^{\frac{m+3}{2}}-11$, $2^{\frac{m+3}{2}}-13$ and $2^{\frac{m+3}{2}}-15$ are coset leaders. Then $(C^{\,(2,n)}_{s_{(i,m)}}\cup C^{\,(2,n)}_{v_{(i,m)}}) \cap A = \emptyset$. By the result (1) of this lemma, we have $\big(B_{(0, i, m)}\backslash (C^{\,(2,n)}_{t_{(i, m)}}\cup C^{\,(2,n)}_{u_{(i, m)}})\big) \cap A = \emptyset$. Therefore,
        \begin{equation*}
            \begin{aligned}
                 &B_{(1,i,m)}\cap A \\
                =&\left(S_{(1\oplus i ,m)}\backslash \left[\big(B_{(0,i,m)}\backslash (C^{\,(2,n)}_{t_{(i,m)}}\cup C^{\,(2,n)}_{u_{(i,m)}})\big)\cup C^{\,(2,n)}_{s_{(i,m)}} 
                \cup C^{\,(2,n)}_{v_{(i,m)}}\right]\right) 
                \cap A\\
                =& S_{(1\oplus i ,m)} \cap A.
            \end{aligned}
        \end{equation*}
        This completes the proof.
    \end{IEEEproof}

    \begin{lemma} 
        \label{Lem5}
        For each $i \in \{0,1\}$, we have 
        \begin{itemize}
            \item [(1)] $\left|T_{(1,i,m)}\right| = 2^{m-1}-1.$
            \item [(2)] $\{j:1\leq j \leq 4\cdot(2^{\frac{m-1}{2}}-4)\} \subseteq T_{(1,i,m)}.$
            \item [(3)] $\{j:0\leq j \leq 2^{\frac{m-1}{2}}-2\} \subseteq \mathbb{Z}_n \backslash -T_{(1,i,m)}.$
        \end{itemize}
    \end{lemma}
    \begin{IEEEproof}
        (1) To prove the desired result, we need to calculate $|A \cup B_{(1, i, m)}|$. By Lemma \ref{Lem4}, we get that $\left|A\cap B_{(1,i,m)}\right| = \left|A\cap S_{(1\oplus i,m)}\right|$. Thus,
        \begin{equation}
            \label{Equation3}
            \begin{aligned}
                &\left|T_{(1,i,m)}\right| \\
                =& \left|A\right| + \left|B_{(1,i,m)}\right| - \left|A\cap B_{(1,i,m)}\right| \\
                =& \left|A\right| + \left|S_{(1\oplus i,m)} \right| 
                - \left|\Big(B_{(0,i,m)}\backslash \big(C_{t_{(i,m)}}^{\,(2,n)}\cup C_{u_{(i,m)}}^{\,(2,n)}\big)\Big) \cup \big(C_{s_{(i,m)}}^{\,(2,n)} \cup C_{v_{(i,m)}}^{\,(2,n)} \big)\right|
                - \left|A\cap S_{(1\oplus i,m)}\right|.
            \end{aligned}
        \end{equation}

        By Lemma \ref{Lem3}, we obtain that $(C^{\,(2,n)}_{s_{(i,m)}}\cup C^{\,(2,n)}_{v_{(i,m)}}) \cap B_{(0,i,m)} =  \emptyset$, then 
        $$\left|\Big(B_{(0,i,m)}\backslash \big(C_{t_{(i,m)}}^{\,(2,n)}\cup C_{u_{(i,m)}}^{\,(2,n)}\big)\Big) \cap \big(C_{s_{(i,m)}}^{\,(2,n)} \cup C_{v_{(i,m)}}^{\,(2,n)} \big)\right| = 0.$$
        By Lemma \ref{Lem1}, we have
        \begin{equation}
            \label{Equation4}
            \begin{aligned}
                &\left|\Big(B_{(0,i,m)}\backslash \big(C_{t_{(i,m)}}^{\,(2,n)}\cup C_{u_{(i,m)}}^{\,(2,n)}\big)\Big) \cup \big(C_{s_{(i,m)}}^{\,(2,n)} \cup C_{v_{(i,m)}}^{\,(2,n)} \big)\right| \\
                =&\left|B_{(0,i,m)}\backslash \big(C_{t_{(i,m)}}^{\,(2,n)}\cup C_{u_{(i,m)}}^{\,(2,n)}\big)\right| + \left|C_{s_{(i,m)}}^{\,(2,n)} \cup C_{v_{(i,m)}}^{\,(2,n)} \right|
                - \left|\Big(B_{(0,i,m)}\backslash \big(C_{t_{(i,m)}}^{\,(2,n)}\cup C_{u_{(i,m)}}^{\,(2,n)}\big)\Big) \cap \big(C_{s_{(i,m)}}^{\,(2,n)} \cup C_{v_{(i,m)}}^{\,(2,n)} \big)\right|\\
                =&\left|B_{(0,i,m)}\right| - \left|C_{t_{(i,m)}}^{\,(2,n)}\cup C_{u_{(i,m)}}^{\,(2,n)}\right| + \left|C_{s_{(i,m)}}^{\,(2,n)} \cup C_{v_{(i,m)}}^{\,(2,n)} \right| \\
                =&\left|B_{(0,i,m)}\right| -2\cdot m + 2\cdot m \\
                =&\left|B_{(0,i,m)}\right|.
            \end{aligned}
        \end{equation}
        For each $i \in \{0,1\}$, let $$A_{(i,m)} = \bigcup_{\substack{j=1\\ wt(j)\equiv i \pmod 2}}^{4\cdot(2^{\frac{m-1}{2}}-4)} C_j^{\,(2,n)}.$$
        Then we have $A_{(i,m)}\cup A_{(i\oplus1,m)} = A$, $A_{(i,m)}\cap A_{(i\oplus1,m)} = \emptyset$ and $A_{(i\oplus1,m)} = A \cap S_{(1\oplus i, m)}$. Note that $C_{n-j_1}^{\,(2,n)} = C_{n-j_2}^{\,(2,n)}$ if and only if $C_{j_1}^{\,(2,n)} = C_{j_2}^{\,(2,n)}$ and $\left|C_{j_1}^{\,(2,n)}\right| = \left|C_{j_2}^{\,(2,n)}\right|$. Then 
        \begin{equation}
            \label{Equation5}
            \begin{aligned}
                \left|A\right| &= \left|A_{(i,m)}\right| + \left|A_{(i\oplus1,m)} \right|
                  = \left|B_{(0,i,m)}\right|+\left|A \cap S_{(1\oplus i, m)}\right|.
            \end{aligned}
        \end{equation}
        Therefore, by Eqs. \eqref{Equation3}, \eqref{Equation4}, and \eqref{Equation5}, $\left| T_{(1,i,m)}\right| = \left|S_{(1\oplus i,m)}\right| = 2^{m-1} -1$. 

        (2) It is easy to see that $\{j:1\leq j \leq 4\cdot(2^{\frac{m-1}{2}}-4)\}\subseteq A \subseteq T_{(1,i,m)}$.

        (3) Since $0\not\in T_{(1,i,m)} $, then $0\in \mathbb{Z}_n \backslash -T_{(1,i,m)}$. To prove $\{j:0\leq j \leq 2^{\frac{m-1}{2}}-2\} \subseteq \mathbb{Z}_n \backslash -T_{(1,i,m)}$, it suffices to demonstrate that $\{n-j:1\leq j \leq 2^{\frac{m-1}{2}}-2\} \subseteq \mathbb{Z}_n \backslash T_{(1,i,m)} $. That is, $\{n-j:1\leq j \leq 2^{\frac{m-1}{2}}-2\} \not \subseteq T_{(1,i,m)} $. It is clear that $\{n-j:1\leq j \leq 2^{\frac{m-1}{2}}-2\}\not \subseteq A$ by Lemma \ref{Lem2}. Then we just need to prove $\{n-j:1\leq j \leq 2^{\frac{m-1}{2}}-2\}\not \subseteq B_{(1,i,m)}$.
    \begin{itemize}
        \item [\romannumeral1)] if $wt(j) = i$, then $n-j \in B_{(0,i,m)}$. Since $n-j \notin A$, it follows from Lemma \ref{Lem4} that $n-j \notin \left(B_{(0,i,m)} \cap A\right) =  C_{t_{(i,m)}}^{\,(2,n)}\cup C_{u_{(i,m)}}^{\,(2,n)}$. Thus, $n-j \in B_{(0,i,m)}\backslash \big(C_{t_{(i,m)}}^{\,(2,n)}\cup C_{u_{(i,m)}}^{\,(2,n)}\big)$. Therefore, we have $n-j \not\in B_{(1,i,m)}$.
        \item [\romannumeral2)] if $wt(j) = 1\oplus i$, then $wt(n-j)=i$ and $n-j\not \in S_{(1\oplus i,m)}$. Note that $B_{(1,i,m)}\subseteq S_{(1\oplus i,m)}$, then we have $n-j \not\in B_{(1,i,m)}$.
    \end{itemize}
    Hence, $\{j:0\leq j \leq 2^{\frac{m-1}{2}}-2\} \subseteq \mathbb{Z}_n \backslash -T_{(1,i,m)}$.
    This completes the proof.
    \end{IEEEproof}
    Now, we are ready to present our new construction of binary cyclic codes.
    \begin{theorem}\label{th6}
        Let $m\geq 9$ be odd. For each $i\in\{0,1\}$, the code $C_{(1,i,m)}$ has parameters
        $$[2^m-1,2^{m-1},d\geq 4\cdot(2^{\frac{m-1}{2}}-4)+1],$$
        and its dual code $C_{(1,i,m)}^\perp$ has parameters
        $[2^m-1,2^{m-1}-1,d^\perp\geq 2^{\frac{m-1}{2}}].$ 
    \end{theorem}
    \begin{IEEEproof}
         It is clear that $\dim(C_{(1,i,m)}) = n - \left|T_{(1,i,m)}\right|$. By Lemma \ref{Lem5}, $\dim(C_{(1,i,m)}) = 2^{m-1}$, hence $\dim(C_{(1,i,m)}^\perp) = n - \dim(C_{(1,i,m)}) = 2^{m-1}-1$.  By (2) and (3) of Lemma \ref{Lem5}, the defining set $T_{(1,i,m)}$ of $C_{(1,i,m)}$  contains the set $\{j:1\leq j \leq 4\cdot(2^{\frac{m-1}{2}}-4)\}$ and the defining set of $C_{(1,i,m)}^\perp$  is $\mathbb{Z}_n \backslash -T_{(1,i,m)}$ which contains the set $\{j:0\leq j \leq 2^{\frac{m-1}{2}}-2\}$. The desired lower bound on $d$ then follows from the BCH bound.
    \end{IEEEproof}

\begin{remark}
    Compared to the binary cyclic codes constructed in \cite{ref5}, which have parameters $[2^m-1,2^{m-1},d\ge3\cdot2^{(m-1)/2}-1]$ and $d^\perp\ge2^{(m-1)/2}$, the cyclic codes presented in Theorem \ref{th6} achieve larger minimum distance while maintaining the same dual distance.
\end{remark}

\begin{remark}
    In this section, we present two families of binary cyclic codes and both the product of $d\cdot d^\perp\approx2n$ when $n$ tends to the infinity. Some specific values of the product of $d\cdot d^\perp$ in recent research have been listed in Tables \ref{table2} and \ref{table1}. Compared with the punctured binary
 Reed-Muller code $R((m-1)/2,m)$, whose $d\cdot d^\perp\approx2n$, the new binary cyclic codes constructed in Sections \ref{V.A} and \ref{V.B} can have minimum distances larger than that of $R((m-1)/2,m)$. Moreover, for the binary cyclic codes with length $n=2^m-1$ and dimension $k=(n\pm1)/2$, except for the punctured binary Reed-Muller codes, few known constructions have reached this bound to the best of our knowledge. 
\end{remark}

\section{Concluding Remarks}\label{set6}
In this paper, we have investigated the binary cyclic code with length $n=2^m-1$ and dimension $k=(n\pm1)/2$. When $m$ is even, we provided a construction of binary $[2^m-1,2^{m-1}-1,d]$ cyclic code with $d\ge2^{m/2}-1$ and $d^\perp\ge2^{m/2}$. When $m$ is odd, we presented two families of binary $[2^m-1,2^{m-1},d]$ cyclic codes with $d\ge2^{(m+1)/2}-1$, $d^\perp\ge2^{(m+1)/2}$ and $d\ge2^{\frac{m+3}{2}}-15$, $d^\perp\ge2^{(m-1)/2}$. When $m$ is the product of two primes in some cases, we constructed some cyclic codes where the dimension is near $\frac{n}{2}$ and $d>\frac{n}{\log_2n}.$

In recent years, much research has been dedicated to constructing suitable defining sets to make the minimum distance $d$ of the binary cyclic code $C$ larger. The known values of $d$ and $d^\perp$ are listed in Tables \ref{table2} and \ref{table1}. However, it seems that the minimum distance $d$ increases while the dual minimum distance $d^\perp$ decreases correspondingly. For the binary cyclic codes with length $n=2^m-1$, when $m$ is even and the product of two distinct primes, we construct a family of cyclic codes whose the value of $d\cdot d^\perp$ reaches $n$ asymptotically in Sections \ref{set00} and \ref{sect4} respectively and when $m$ is odd. In addition, in Section \ref{set3}, we present two families of binary cyclic codes with $d\cdot d^\perp\ge 2n$ asymptotically. To the best of our knowledge, few constructions of binary cyclic codes have reached this bound. It remains an interesting problem to construct more infinite families of cyclic codes of length $n$ whose $d\cdot d^\perp$ has a larger lower bound.
 
Furthermore, denote $\Delta=\limsup\limits_{n \rightarrow \infty}\frac{d \cdot d^{\perp}}{n}$, which represents the largest possible value of $\frac{d \cdot d^{\perp}}{n}$ for any family of binary cyclic codes with dimension $k$ near $\frac{n}{2}$, minimum distance $d$, dual distance $d^{\perp}$ and code length $n$ tending to infinity. The punctured binary
 Reed-Muller codes $R((m-1)/2,m)$ and two families of binary cyclic codes constructed in Section \ref{set3} show that $\Delta \geq 2$. Therefore, we propose the following open problem.

\textbf{Open Problem :}  
\begin{itemize}
    \item Is $\Delta >2?$ Equivalently, does there exist an infinite family of binary cyclic codes with dimension $k$ near $\frac{n}{2}$, minimum distance $d$, dual distance $d^{\perp}$ and code length $n$, such that $\limsup\limits_{n \rightarrow \infty}\frac{d \cdot d^{\perp}}{n}>2$?
    \item Determine the exact value of $\Delta$.
\end{itemize}

\section*{Acknowledgement}
The authors sincerely thank the Associate Editor, Prof. Ferruh Ozbudak, and the anonymous reviewers for their valuable
comments, which have significantly improved the manuscript.

\appendices

\section{Proof of Theorem \ref{thm6}}\label{B}
\begin{IEEEproof}
    Let $D_0=\left\{1,2,...,\left\lfloor\frac{2^{p_1p_2}-1}{p_1p_2}\right\rfloor,\left\lfloor\frac{2^{p_1p_2}-1}{p_1p_2}\right\rfloor+1,...,\left\lfloor\frac{2^{p_1p_2}-1}{p_1p_2}\right\rfloor+\left\lfloor\frac{2^{p_1}-1}{2p_1p_2}\right\rfloor+\left\lfloor\frac{2^{p_2}-1}{2p_1p_2}\right\rfloor+\left\lfloor\frac{2^{p_1}-1}{4p_1p_2}\right\rfloor+\left\lfloor\frac{2^{p_2}-1}{4p_1p_2}\right\rfloor-1\right\}$ given in the proof of Theorem \ref{thm5}. Since
    \begin{align*}
        &\frac{2^{p_1p_2+1}}{2^{\lfloor\log_22p_1p_2\rfloor}}-\left\lfloor\frac{2^{p_1p_2}-1}{p_1p_2}\right\rfloor-\left\lfloor\frac{2^{p_1}-1}{2p_1p_2}\right\rfloor-\left\lfloor\frac{2^{p_2}-1}{2p_1p_2}\right\rfloor
        -\left\lfloor\frac{2^{p_1}-1}{4p_1p_2}\right\rfloor-\left\lfloor\frac{2^{p_2}-1}{4p_1p_2}\right\rfloor\\
        \ge&\frac{2^{p_1p_2+1}}{2p_1p_2-1}-\frac{2^{p_1p_2}-1}{p_1p_2}-\frac{3(2^{p_2}-1)}{4p_1p_2}-\frac{3(2^{p_1}-1)}{4p_1p_2}\\
        =&\frac{2^{p_1p_2}+2p_1p_2-1}{(2p_1p_2-1)p_1p_2}-\frac{3(2^{p_2}-1)}{4p_1p_2}-\frac{3(2^{p_1}-1)}{4p_1p_2}\\
        >&0,
    \end{align*}
 then for any $1\le i\le\left\lfloor\frac{2^{p_1p_2}-1}{p_1p_2}\right\rfloor+\left\lfloor\frac{2^{p_1}-1}{2p_1p_2}\right\rfloor+\left\lfloor\frac{2^{p_2}-1}{2p_1p_2}\right\rfloor+\left\lfloor\frac{2^{p_1}-1}{4p_1p_2}\right\rfloor+\left\lfloor\frac{2^{p_2}-1}{4p_1p_2}\right\rfloor-1$, we have $i\le2^{p_1p_2+1-\lfloor\log_22p_1p_2\rfloor}-2$, which leads to $wt(\pi(i))\le p_1p_2-\lfloor\log_22p_1p_2\rfloor$. Thus, by Corollary \ref{cor1}, for any $x\in\cup_{x\in D_0}C^{(2,n)}_x$, $wt(\pi(x))\le p_1p_2-\lfloor\log_22p_1p_2\rfloor.$

 Note that for any $1\le j\le2^{\lfloor\log_2p_1p_2\rfloor+1}-1$, we have $wt(\pi(2^{p_1p_2}-j))\ge p_1p_2-\lfloor\log_2p_1p_2\rfloor>p_1p_2-\lfloor\log_22p_1p_2\rfloor$, which leads to $2^{p_1p_2}-j\not\in\cup_{x\in D_0}C_x^{(2,n)}$. Furthermore, since $\pi(2^{p_1p_2}-j)=(\bm{1}_{p_1p_2-\lfloor\log_2p_1p_2\rfloor-1},\bm{a})$ with some $\bm{a}\in\mathbb{F}_2^{\lfloor\log_2p_1p_2\rfloor+1}$ then by Corollary \ref{cor3}, for any $x\in D_0$, we have $\left|C_x^{(2,n)}\right|=p_1p_2$ since $p_1p_2-\lfloor\log_2p_1p_2\rfloor-1>p_2>p_1$.

If $\frac{2^{p_1p_2}-2^{p_1}-2^{p_2+1}+4}{2p_1p_2}\le\sigma_0<\frac{2^{p_1p_2}-2^{p_1}}{2p_1p_2}$, from Eq. \eqref{9}, we have $\left|C_x^{(2,n)}\right|\neq p_1p_2$ for any $x\in Z_2(\sigma_0,1,2,...,\sigma_0)\backslash(\bigcup_{x\in D_0}C_x^{(2,n)})$. Thus $2^{p_1p_2}-j\notin Z_1(\sigma_0;1,2,...,\sigma_0)\backslash(\bigcup_{x\in D_0}C_x^{(2,n)})$, hence $2^{p_1p_2}-j\not\in Z_2(\sigma_0;1,2,...,\sigma_0)$.

      If $\sigma_0<\frac{2^{p_1p_2}-2^{p_1}-2^{p_2+1}+4}{2p_1p_2}$,  note that  
      \begin{align*}
          \left|\left\{2^{p_1p_2}-j: j=2,...,2^{\lfloor\log_2p_1p_2\rfloor+1}-1\right\}_{odd}\right|=2^{\lfloor\log_2p_1p_2\rfloor}-1<\frac{2^{p_1p_2}-2^{p_1}-2^{p_2}+2}{p_1p_2}-\frac{2^{p_1p_2}-2^{p_1}-2^{p_2+1}+4}{2p_1p_2},
      \end{align*}
then we can always choose suitable $\sigma_0<\gamma_{\sigma_0+1}<\gamma_{\sigma_0+2}<\cdots<\gamma_{\frac{2^{p_1p_2}-2^{p_1}-2^{p_2+1}+4}{2p_1p_2}}\leq \frac{2^{p_1p_2}-2^{p_1}-2^{p_2}+2}{p_1p_2}$ such that $2^{p_1p_2}-j\not\in\bigcup_{l=\sigma_0+1}^{\frac{2^{p_1p_2}-2^{p_1}-2^{p_2+1}+4}{2p_1p_2}}C_{k_{\gamma_l}}^{(2,n)}$. Thus, for any $1\le j\le2^{\lfloor\log_2p_1p_2\rfloor+1}-1$, 
      $$2^{p_1p_2}-j\not\in\left(\bigcup_{\gamma=1}^{\sigma_0}C_{k_{\gamma}}^{(2,n)}\right)\bigcup\left(\bigcup_{l=\sigma_0+1}^{\frac{2^{p_1p_2}-2^{p_1}-2^{p_2+1}+4}{2p_1p_2}}C_{k_{\gamma_l}}^{(2,n)}\right).$$ 

   In summary, for any $2^{p_1p_2}-2^{\lfloor\log_2p_1p_2\rfloor+1}+1\le x\le 2^{p_1p_2}-1$, we always have $x\not\in Z_2(\sigma;\gamma_1,\gamma_2,...,\gamma_\sigma)$, which leads to $\left\{0,1,...,2^{\lfloor\log_2p_1p_2\rfloor+1}-2\right\}\subseteq-(\mathbb{Z}_n\backslash Z_2(\sigma;\gamma_1,\gamma_2,...,\gamma_\sigma)).$ By the BCH bound, $d^\perp\ge2^{\lfloor\log_2p_1p_2\rfloor+1}.$
\end{IEEEproof}
    
\bibliographystyle{IEEEtran}
\bibliography{bibliofile}
\end{document}